\documentclass[10pt]{article}
\usepackage [top=25truemm,bottom=25truemm,left=25truemm,right=25truemm]{geometry}
\usepackage{fancybox}
\usepackage{latexsym,amssymb,amsfonts}
\usepackage[dvips]{graphicx}
\usepackage{cite}
\usepackage{mysty}
\usepackage{yhmath}
\DeclareMathOperator*{\wlim}{wlim}
\DeclareMathOperator*{\supp}{supp}
\DeclareMathOperator*{\sgn}{sgn}

\DeclareMathOperator*{\im}{Im}

\newcommand{\rL}{{\rm L}}
\newcommand{\rR}{{\rm R}}
\newcommand{\rA}{{\rm A}}
\newcommand{\rG}{{\rm G}}
\newcommand{\rP}{{\rm P}}
\newcommand{\rC}{{\rm C}}
\usepackage{subcaption}

\def\revise{}
\title{\bf Skeleton structure inherent\\in discrete-time quantum walks}

\author{Tomoki Yamagami $^{1,\,*}$
\and Etsuo Segawa $^2$
\and Ken'ichiro Tanaka $^3$
\and Takatomo Mihana $^1$
\and Andr\'e R\"ohm $^1$
\and Ryoichi Horisaki $^1$
\and Makoto Naruse $^1$
}
\date{}

\begin{document}
\maketitle
\vspace{-2.1\baselineskip}
\begin{center}
{\small
$^1$ Department of Information Physics and Computing, Graduate School of Information Science and Technology,\\
The University of Tokyo, 7-3-1 Hongo, Bunkyo, Tokyo 113-8656, Japan.\\
$^2$ Graduate School of Environment and Information Sciences, Yokohama National University,\\
79-1 Tokiwadai, Hodogaya, Yokohama, Kanagawa 240-8501, Japan.\\
$^3$ Department of Mathematical Informatics, Graduate School of Information Science and Technology,\\
The University of Tokyo, 7-3-1 Hongo, Bunkyo, Tokyo 113-8656, Japan\\
$^*$ Corresponding author. Email: \texttt{yamagami-tomoki-qwb@g.ecc.u-tokyo.ac.jp}
}\vspace{1\baselineskip}\\
\end{center}

\abstract
In this paper, we \revise{claim} that a common underlying structure---a skeleton structure---is present behind discrete-time quantum walks \revise{(QWs)} on a one-dimensional lattice with a homogeneous coin matrix. 
\revise{This skeleton structure is independent of the initial state, and partially, even of the coin matrix.}
\revise{This structure is best interpreted in the context of quantum-walk-replicating random walks (QWRWs), i.e., random walks that replicate the probability distribution of quantum walks, where this newly found structure acts as a simplified formula for the transition probability.}
\revise{Additionally,} we \revise{construct} a random walk whose transition probabilities are defined by the skeleton structure \revise{and} demonstrate that the resultant properties of the walkers are similar to \revise{both} the original \revise{QWs and QWRWs.}

\keyword
Quantum walk, Random walk, Quantum-walk-replicating random walk, Transition probability

\section{Introduction}\label{sec:intro}
A quantum walk (QW) is the quantum counterpart of the classical random walk \cite{NK08,JK03,SEV12,VK07}, which it extends by including the effects of quantum superposition and complex probability amplitudes. 
QWs were first introduced in the field of quantum information theory \cite{SG88,YA93,AA01}. 
The characteristic structure of quantum walks was intensively studied by mathematicians \cite{NK02,NK05}, and since then, quantum walks have been important subjects in both fundamental and applied research. 
Indeed, quantum walks exhibit varying behavior depending on the conditions or settings of time and space, so there have been many studies on the mathematical analysis of these evolution models \cite{NK10, TS12, NK13a, JB14, CG19, CC20}. 
In addition, their unique behavior is useful for implementing quantum structures or quantum analogs of existing models; therefore, their application has been considered in fields such as quantum teleportation \cite{TY21,YW17}, time series analysis \cite{NK19}, topological insulators \cite{JA13, HO15}, radioactive waste reduction \cite{LM11, AI15}, and optics \cite{YI17}.

The main properties of QWs are \textit{linear spreading} and \textit{localization}. 
The former means that the standard deviation of the distribution of the measurement probability of quantum walkers (QWers) grows in proportion to the run time $n\in\mathbb{N}_0  := \mathbb{N}\cup \{0\}$. 
The latter implies that the probability is distributed at a particular position no matter how long the walk runs. 
QWs have both or either of these properties, resulting in a probability distribution that is totally different from those of random walks, which weakly converge to normal distributions.

\revise{In this paper, we newly introduce an underlying structure of quantum walks which we call the skeleton structure. The skeleton structure is based on the quantities denoted by $p_n(x)$ and $q_n(x)$, which appear when describing the quantum walk in a certain way, and encapsulates how the properties of quantum walks arise. 
The quantities $p_n(x)$ and $q_n(x)$ are interpreted in the context of random walks whose probability distribution is identical to that of one-dimensional discrete-time quantum walks (quantum-walk-replicating random walks; QWRWs) \cite{TY22a}; they play the roles of the transition probabilities in the left and right directions, respectively. 
The transition probabilities determine the trajectories of walkers that are affected by the properties of quantum walks as mentioned above, and in our previous paper \cite{TY22a}, we obtained new insights into quantum walks through the trajectories.}

\revise{The quantities $p_n(x)$ and $q_n(x)$, however, possess a highly complex shape because they have fluctuations inside the positions where the probability distribution is the highest, and thus it has been difficult to quantitatively explain the behavior of quantum walks with these quantities.
By considering the weak limit that excludes the oscillatory behavior of these new terms, $p_n(x)$ and $q_n(x)$, we obtain the skeleton structure in the form of quite a simple formulae. Our study focuses on the case of a homogeneous coin, where only linear spreading is observed. In this case, it is noteworthy that the skeleton structure does \textit{not} depend on the initial state nor, partially, on the coin matrix of the quantum walk. This is despite the fact that the full description of $p_n(x)$ and $q_n(x)$, including oscillatory terms, does significantly depend on the coin matrix and the initial conditions. Analyzing this structure allows the visualization of the behavior that the walkers perform.
}

\revise{To additionally investigate the implication of the skeleton structure,} we define a new time- and site-dependent random walk whose transition probabilities are directly defined by the skeleton structure\revise{, which we call the \textit{quantum-skeleton random walk} (QSRW).}
This structure is robbed of its oscillation, meaning that the resulting random walk is a simplified model of the QWRW. 
Furthermore, we compare the new random walk with the QWRW and show that it still has linear spreading. 

The rest of this paper is organized as follows. First, we review quantum walks (QWs) in Section~\ref{sec:qw}. 
\revise{Then, we formally} define the skeleton structure of QWs and present its underlying statements as our main results in Section~\ref{sec:ss}.
\revise{Here, we introduce the QWRW to give an interpretation of the skeleton structure.} 
In Section~\ref{sec:qsrw}, we construct the QSRW, whose transition probabilities are defined by the skeleton structure, and compare this with QWRWs. Section~\ref{sec:proof} is devoted to the proofs of the theorems presented in Sec.~\ref{sec:ss}. Finally, we give a summary and discussion in Section~\ref{sec:summary}.


\section{Quantum walk (QW)}\label{sec:qw}\setcounter{equation}{0}
We introduce discrete-time quantum walks on a one-dimensional lattice ($\mathbb{Z}$). 
The space of QWs is defined in a compound Hilbert space consisting of the position Hilbert space $\mathcal{H}_{\rP} = \mathrm{span}\{ \ket{x}\,|\,x\in \mathbb{Z}\}$ and the coin Hilbert space $\mathcal{H}_{\rC} = \mathrm{span}\{ \ket{\rL},\,\ket{\rR}\}$ with $\ket{\rL} = [1\,\,\,0]^\trp$ and $\ket{\rR} = [0\,\,\,1]^\trp$. \revise{Here, for a matrix $M$, $M^\trp$ represents its transpose.} Note that \revise{$\mathcal{H}_{\rP}$ and} $\mathcal{H}_{\rC}$ \revise{are} equivalent to \revise{
\begin{align}\label{eq:l2z}
    \ell^2(\mathbb{Z}) := \left\{f:\mathbb{Z}\to \mathbb{C}\,\middle|\,\sum_{x\in\mathbb{Z}}|f(x)|^2< \infty\right\}
\end{align}
and} $\mathbb{C}^2$\revise{, respectively}. Then, the whole system is described by
\begin{align}\label{eq:hilbert}
    \revise{\mathcal{H} =\mathcal{H}_{\rP}\otimes \mathcal{H}_{\rC}}
    \mathrel{\revise{:=\,}} \revise{\mathrm{span}\{\ket{x}\otimes \ket{\mathrm{J}}\,|\, x\in\mathbb{Z},\ \mathrm{J} = \rL,\,\rR\}.}
\end{align}

We consider the state of QWs as follows: for $n\in\mathbb{N}_0$,
\begin{align}\label{eq:state}
	\ket{\varPsi_n} = \sum_{x\in\mathbb{Z}}\ket{x}\otimes \ket{\varPsi_n(x)} \in\mathcal{H}.
\end{align}
Here, $n\in\mathbb{N}_0$ represents run time of QWs, and $\ket{\varPsi_n(x)}\in\mathbb{C}^2$ is called the probability amplitude vector on the position $x\in\mathbb{Z}$ at run time $n$. 
Besides, we set the initial state as
\begin{align}\label{eq:init}
	\ket{\varPsi_0} = \ket{0}\otimes \ket{\varphi}, 
\end{align}
where $\ket{\varphi}\in\mathbb{C}^2$ is a constant vector with $\|\varphi\|=1$. In other words, it follows that
\begin{align}\label{eq:inittqo}
	\ket{\varPsi_0(x)} = \mathbf{1}_0(x)\ket{\varphi},
\end{align}
\revise{where, for $a\in\mathbb{R}$,}
\begin{align}\label{eq:ind}
    \revise{\mathbf{1}_a(x) = \left\{\begin{array}{ll} 1 & (x=a) \\ 0 & (x\not =a)\end{array}\right..}
\end{align}
The time evolution of the system is defined as follows:
\begin{align}\label{eq:tev}
	\ket{\varPsi_{n+1}} = U\ket{\varPsi_n}.
\end{align}

Now, we introduce the unitary operator $U$ to describe the time evolution of the system of QWs: for run time $n\in\mathbb{N}_0$,
\begin{align}\label{eq:U}
U = \bfit{S}\bfit{C}, 
\end{align}
which are defined with a shift operator $\bfit{S}$ and a coin operator $\bfit{C}$ as follows:
\begin{align}\label{eq:shift}
\bfit{S} = S^{-1} \otimes \ketbra{\rL}{\rL} + S \otimes \ketbra{\rR}{\rR} 
\end{align}
with
\begin{align}\label{eq:Shift}
S = \sum_{x \in \mathbb{Z}} \ketbra{x+1}{x},
\end{align}
and
\begin{align}\label{eq:Coin}
\bfit{C} = \sum_{x\in\mathbb{Z}}\ketbra{x}{x} \otimes C. 
\end{align}
Here, $S$ represents the transition from position $x$ to $x+1$ (right direction), and $S^{-1} = \sum_{x\in\mathbb{Z}}\ketbra{x-1}{x}$ indicates the transition from position $x$ to $x-1$ (left direction). 
$C$ is a unitary matrix called a {\it coin matrix}. 
In particular, since $C$ is independent of time $n$ and position $x$, $C$ is called a \textit{homogeneous coin matrix}. We can describe $C$ with \revise{$\delta\in[-\pi,\,\pi)$ and $a,\ b\in \mathbb{C}$ satisfying $|a|^2 + |b|^2 =1$} as
\begin{align}\label{eq:coin}
C=\twobytwo{a}{b}{-e^{i\delta}\overline{b}}{e^{i\delta}\overline{a}},
\end{align}
without loss of generality because of unitarity.

Here, by applying the property of the Kronecker product, we have
\begin{align}\label{eq:UPQ}
	U = S^{-1} \otimes P + S\otimes Q,
\end{align}
where
\begin{align}\label{eq:decompp}
	P = \ketbra{\rL}{\rL}C = \twobytwo{a}{b}{0}{0}
\end{align}
and
\begin{align}\label{eq:decompq}
    Q = \ketbra{\rR}{\rR}C = \twobytwo{0}{0}{-e^{i\delta}\overline{b}}{e^{i\delta}\overline{a}}. 
\end{align}
The matrices $P$ and $Q$ are considered to be the decomposition elements of $C$; that is, the relation $P+Q=C$  holds. 
This decomposition is called the Ambainis type \cite{AA01}; there exists another type of decomposition (see Appendix~\ref{app:tf}).

By Eqs.~\refeq{eq:tev} and \refeq{eq:UPQ}, we have
\begin{align}\label{eq:tevPQ}
	\ket{\varPsi_{n+1}(x)} = P\ket{\varPsi_n(x+1)} +Q\ket{\varPsi_n(x-1)}. 
\end{align}
Moreover, from the initial state \refeq{eq:init}, there exists $\varXi_n(x)$ such that
\begin{align}\label{eq:state_xi}
	\ket{\varPsi_n(x)} = \varXi_n(x)\ket{\varphi}.
\end{align}
$\varXi_n(x)$ describes the weight of all the possible paths from the origin to the position $x$ at run time $n$, which can be written as
\begin{align}\label{eq:varXi}
\varXi_n(x) = \sum_{(\ell_j,\,r_j)\in \mathcal{P}}\left(\prod_{j=1}^{n}P^{\ell_j}Q^{r_j}\right), 
\end{align}
where
\begin{align}\label{eq:scope}
\mathcal{P} =\Biggl\{&\{(\ell_j,\,r_j)\}_{j=1}^{n}\in \{(1,\,0),\,(0,\,1)\}^n\, 
\Biggl|\, \Biggl(\sum_{j=1}^{n}\ell_j =\fraction{n-x}{2}\Biggr) \land \Biggl(\sum_{j=1}^{n}r_j =\fraction{n+x}{2}\Biggr)\Biggr\}.
\end{align}
Moreover, from Eq.~\refeq{eq:tevPQ}, the following relation holds:
\begin{align}\label{eq:tevXi}
	\varXi_{n+1}(x) = P\varXi_n(x+1) + Q\varXi_n(x-1). 
\end{align}

At the end of the definition of QWs, the measurement probability of the particle at position $x$ at run time $n$, denoted by $\mu_n(x)$, is given by
\begin{align}\label{eq:measurement}
	\mu_n(x) = \mathbb{P}(X_n =x) := \|\varPsi_n(x)\|^2,
\end{align}
where $X_n$ is the position of a walker that follows a QW, called a QWer, at time $n$, \revise{and $\mathbb{P}$ donates a probability.} This is the application of Born's rule of quantum mechanics. Note that for any $n\in\mathbb{N}_0$, the following is satisfied:
\begin{align}\label{eq:measurementisprobmeasure}
\sum_{x\in\mathbb{Z}}\mu_n(x) = \sum_{x\in\mathbb{Z}}\|\varPsi_n(x)\|^2 = \|\varPsi_n\|^2 = 1.
\end{align}

\section{Skeleton structure of quantum walks}\label{sec:ss}\setcounter{equation}{0}
{In this section, as our main results, we introduce a quantity representing a property of quantum walks, which we call the \textit{skeleton structure} of quantum walks. This structure is independent of the initial state and, partially, even the coin matrix.}

{First of all, for the pair $(n,\,x)\in \mathbb{N}_0\times \mathbb{Z}$, we consider the functions $p_n(x)$ and $q_n(x)$ satisfying the following:
\begin{align}
	&\|P\varPsi_n(x)\|^2 = p_n(x)\|\varPsi_n(x)\|^2, \label{eq:pnx0}\\
	&\|Q\varPsi_n(x)\|^2 = q_n(x)\|\varPsi_n(x)\|^2. \label{eq:qnx0}
\end{align}
Note that $p_n(x)$ and $q_n(x)$ are arbitrary if the relation $\|\varPsi_n(x)\|^2 = 0$ holds. If this relation does not hold, we can rewrite Eqs.~\refeq{eq:pnx0} and \refeq{eq:qnx0} as
\begin{align}
	&p_n(x) = \fraction{\|P\varPsi_n(x)\|^2}{\|\varPsi_n(x)\|^2}, \label{eq:pnx1}\\
	&q_n(x) = \fraction{\|Q\varPsi_n(x)\|^2}{\|\varPsi_n(x)\|^2}. \label{eq:qnx1}
\end{align}
It is worth remarking that, by the definitions of the matrices $P$ and $Q$ (Eqs.~\refeq{eq:decompp} and \refeq{eq:decompq}), the relation
\begin{align}\label{eq:vPsi}
	\|P\varPsi_n(x)\|^2 + \|Q\varPsi_n(x)\|^2 = \|\varPsi_n(x)\|^2
\end{align}
holds; that is, $p_n(x) + q_n(x) = 1$. From this perspective, we can consider that $p_n(x)$ and $q_n(x)$ are the ratios how the values of $\|\varPsi_n(x)\|^2$ are distributed to $\|P\varPsi_n(x)\|^2$ and $\|Q\varPsi_n(x)\|^2$, respectively.
}

{The functions $p_n(x)$ and $q_n(x)$ can be interpreted as the transition probabilities of random walks that replicate a probability distribution of quantum walks. Before defining the skeleton structure, we introduce the quantum-walk-replicating random walk (QWRW) \cite{TY22a}.}
{\subsection{Quantum-walk-replicating random walk (QWRW)}\label{subsec:qwrw}}
\begin{table}[b]
\begin{center}
\begin{tabular}{|c||c|c|}\hline
	 & Symmetric case\vrule width 0pt height 10pt depth 0pt& Asymmetric case\\ \hline\hline
$a$ & $1/\sqrt{2}$\vrule width 0pt height 12pt depth 5pt& $1/\sqrt{3}$\\ \hline
$b$ & $1/\sqrt{2}$\vrule width 0pt height 12pt depth 5pt& ${\sqrt{2/3}}$\\ \hline
$\delta$ & $\pi$\vrule width 0pt height 10pt depth 0pt& $0$\\ \hline
$\ket{\varphi}$ & $[1\,\,\,i]^\trp/\sqrt{2}$\vrule width 0pt height 12pt depth 5pt& $[1\,\,\,0]^\trp$\\ \hline
\end{tabular}
\caption{Parameters of the symmetric and asymmetric cases.}\label{tb:cases}
\end{center}
\end{table}

\begin{figure*}[t] 
\includegraphics[width=\linewidth]{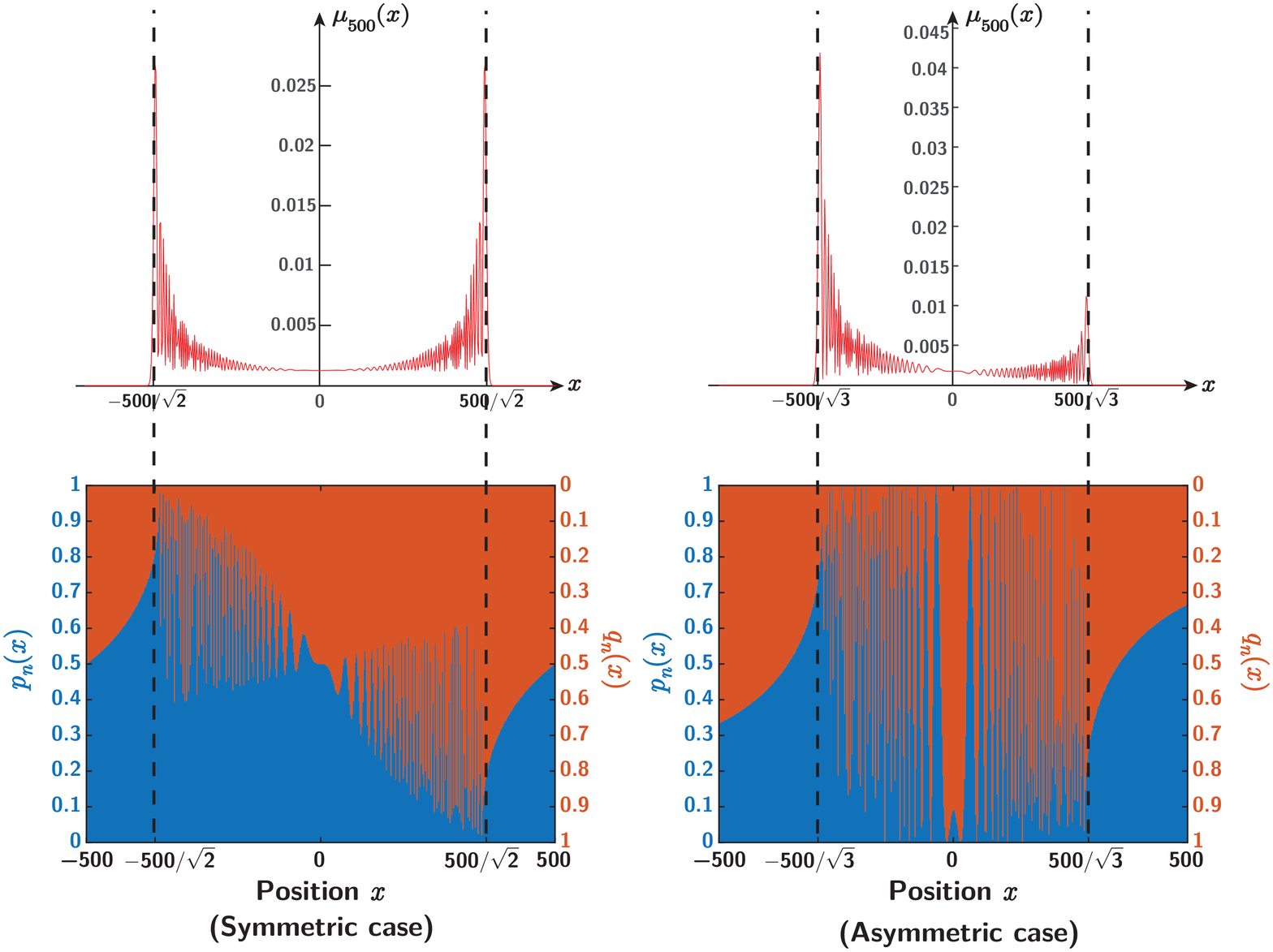}
\caption{
\revise{Probability distributions $\mu_n(x)$ of QW or QWRW} and functions $p_n(x)$ and $q_n(x)$ in the symmetric and asymmetric cases \revise{at time steps $n=500$}. 
The parameters in both cases are summarized in Table~\ref{tb:cases}. 
Each value is plotted only on even points; on odd points, the probabilities are $0$. 
\revise{Functions $p_{500}(x)$ and $q_{500}(x)$ on the lower panel} are expressed \revise{with} a stacked bar graph, which is drawn only on the even points. Note that the relation $p_{500}(x) + q_{500}(x) = 1$ holds for all $x$. {The vertical dotted lines indicate the position of the peaks of $\mu_{500}$ at $x=\pm 500|a|$.}
}\label{fig:pq}
\vrule height 0.3mm width 165mm
\end{figure*}

{By using the time evolution of the QWs (Eq.~\refeq{eq:tevPQ}) and the definitions of the matrices $P$ and $Q$ (Eqs.~\refeq{eq:decompp} and \refeq{eq:decompq}), we obtain the following relation:
\begin{align}\label{eq:rec_norm_of_varPsi0}
	\|\varPsi_{n+1}(x)\|^2 = \|P\varPsi_n(x+1)\|^2 + \|Q\varPsi_n(x-1)\|^2.
\end{align}
By using Eqs.~\refeq{eq:pnx0} and \refeq{eq:qnx0}, we can construct the recurrent formula of $\|\varPsi_n(x)\|^2$ with $p_n(x)$ and $q_n(x)$ as follows:
\begin{align}\label{eq:rec_norm_of_varPsi1}
	\|\varPsi_{n+1}(x)\|^2 = p_n(x+1)\|\varPsi_n(x+1)\|^2 + q_n(x-1)\|\varPsi_n(x-1)\|^2.
\end{align}
Here we recall that $\|\varPsi_n(x)\|^2$ represents the measurement probability $\mu_n(x)$ at position $x$ at time step $n$ (see \refeq{eq:measurement}). Thus, we obtain
\begin{align}\label{eq:rec_mu}
	\mu_{n+1}(x) = p_n(x+1)\mu_n(x+1) + q_n(x-1)\mu_n(x-1).
\end{align}
This formula describes the main idea behind the QWRW. As described in our previous work \cite{TY22a}, a random walk that replicates the probability distribution of a quantum walk can be constructed with the help of $p_n(x)$ and $q_n(x)$, where these terms become the site- and time-dependent transition probabilities to the left and right sides, respectively. These transition probabilities are formally determined as
\begin{align}\label{eq:p}\begin{split}
	&\mathbb{P}(S_{n+1} = x-1\,|\,S_n = x) := p_n(x)\\
	&\hspace{0.1\textwidth}= \left\{\begin{array}{cl}\fraction{\|P\varPsi_n(x)\|^2}{\mu_n(x)} & (\mu_n(x) > 0)\vrule width 0pt height 0pt depth 15pt\\ \fraction{1}{2} & (\mu_n(x) = 0) \end{array}\right., 
\end{split}\end{align}
\begin{align}\label{eq:q}\begin{split}
	&\mathbb{P}(S_{n+1} = x+1\,|\,S_n = x) := q_n(x)\\
	&\hspace{0.1\textwidth}= \left\{\begin{array}{cl}\fraction{\|Q\varPsi_n(x)\|^2}{\mu_n(x)}& (\mu_n(x) > 0)\vrule width 0pt height 0pt depth 15pt\\ \fraction{1}{2} & (\mu_n(x) = 0) \end{array}\right., 
\end{split}\end{align}
}where $S_n$ is the position of a QWRWer at time $n$. It should be noted that {defining $p_n(x)$ and $q_n(x)$ for a pair $(n,\,x)$ which satisfies $\mu_n(x)=0$ is only a formal treatment; recall that we can define them arbitrarily.} {Note that} $p_n(x)$ and $q_n(x)$ satisfy the necessary condition to be transition probabilities: they take real values between $0$ and $1$, and $p_n(x) + q_n(x) = 1$ for all pairs $(n,\,x)\in \mathbb{N}_0\times \mathbb{Z}$.

{Fig.~\ref{fig:pq} shows the numerical results of the calculations of $p_n(x)$ and $q_n(x)$ and their relationship to the probability distributions of QWRWs, which are naturally identical to those of the corresponding quantum walks; a symmetric case and an asymmetric case have the parameters $(a,\,b,\,\delta,\,\ket{\varphi})$ as shown in Table~\ref{tb:cases}, respectively. Note that the parameters $(a,\,b,\,\delta)$ in the symmetric case construct a coin matrix called the Hadamard matrix.
A common property of these two cases is that the graphs of $p_n(x)$ and $q_n(x)$ can be divided into two kinds of regions: oscillatory and non-oscillatory regions. The boundaries of these parts correspond to the peaks of the probability distribution of the QWs, which are at peaks $x=\pm n|a|$ at run time $n$.}

We can regard $p_n(x)$ and $q_n(x)$ as a kind of ``mean velocity" of walkers in the left and right directions, respectively. The larger $p_n(x)$ becomes, the faster the walkers go to the left side, and the larger $q_n(x)$ is, the faster the walkers go to the right side. The transition probabilities have \revise{high-frequency} oscillations inside the peaks, which make them difficult to analyze. However, we can capture their ``tendency." Inside the peaks, the farther from the origin a walker is, the faster it tends to move away from the origin; this corresponds to the property that the \revise{local maxima} of $p_n(s)$ on the left side is larger than that on the right side, and vice versa. Outside the peaks, this relation is reversed; that is, the farther away from the origin a walker is, the slower it moves away from it. This is caused by the change of the slope of $p_n$ and $q_n$ at $x=\pm n|a|$. We can infer that the behavior on both sides of the peaks causes walkers to accumulate around $x=\pm n|a|$ and say that the functions $p_n$ and $q_n$ indicates how the peaks of QW-distributions are made through the interpretation of QWRWs.

The structure of $p_n(x)$ and $q_n(x)$ as seen in Fig.~\ref{fig:pq} is quite complicated and not easy to interpret. 
Not only do $p_n(x)$ and $q_n(x)$ strongly depend on the spatial position $x$ inside the peaks, but there are also temporal oscillations with $n$. \revise{These two kinds of oscillations prevent us to clearly explain the tendency mentioned above (see Appendix~\ref{app:oscillation} regarding the detailed discussions of oscillations).}

The strong fluctuations of $p_n$ and $q_n$ make it difficult to quantitatively explain the behavior of quantum walks through QWRWs. This is one of the reasons why we want to consider the skeleton structure of quantum walks as the simplified model thereof.

\subsection{Main theorems}\label{subsec:statements}
We present our main theorems from which we construct the skeleton structure of quantum walks and its relation to $p_n(x)$ and $q_n(x)$ as described above.
Detailed proofs are given later in Section~\ref{sec:proof}. These are based on the original calculations using the original quantum walk and its time evolution.

Inside the peaks, oscillations do not decay with time $n$, and thus $p_n(x)$ does not converge with increasing $n$. 
{Nevertheless, we} can characterize the transition probabilities in the following form with the notion of a weak limit. 

\begin{thm}[\textbf{Inside the peaks}]\label{thm:inside}
Let $x_n \in\mathbb{Z}$ and $s \in (-|a|,\,|a|)$ satisfy
\begin{align}\label{eq:xn1}
	x_n = ns + O\left(\fraction{1}{n}\right)\ \text{as $n\to \infty$}.
\end{align}
Then, $p_n(x_n)$ and $q_n(x_n)$ can be described as
\begin{align}
	&p_n(x_n) = \fraction{\tau_1(s) + \xi_n(s)}{1 + \eta_n(s)}, \label{eq:pnxn}\\
	&q_n(x_n) = \fraction{(1-\tau_1(s))+ \zeta_n(s)}{1 + \eta_n(s)},\label{eq:qnxn}
\end{align}
where
\begin{align}
	&\tau_1(s) = \fraction{1-s}{2},\label{eq:tauinside}\\ &\wlim_{n\to\infty}\xi_n(s)=\wlim_{n\to\infty}\eta_n(s)=\wlim_{n\to\infty}\zeta_n(s) = 0. \label{eq:wlimit}
\end{align}
Here, for the sequence $\{f_n\}_{n\in\mathbb{N}_0}\subset \mathbf{L}^2(\mathbb{R})$, there exists $f\in\mathbf{L}^2(\mathbb{R})$ such that $\wlim_{n\to\infty}f_n(s) = f(s)$ ($f_n(s)$ weakly converges to $f(s)$) iff for any function $g (s)\in\mathbf{L}^2(\mathbb{R})$ and $y\in\mathbb{R}$,
\begin{align}\label{eq:wl}
	\lim_{n\rightarrow\infty}\int_{-\infty}^{y}f_n(s)g(s)\,\rd s = \int_{-\infty}^{y}f(s)g(s)\,\rd s.
\end{align}
\end{thm}

The forms represented in Theorem~\ref{thm:inside} are the decomposition of the numerator and denominator of $p_n(x_n)$ and $q_n(x_n)$ into the terms dependent on or independent of run time $n$. 

The functions $\xi_n$, $\eta_n$ and $\zeta_n$ correspond to the former terms; they include a term that oscillates and one that decays with the growth of $n$ on the order of $1/n$. 
Although these terms do not converge to specific values, weak convergence holds. 

In contrast, the function $\tau_1(s)$ corresponds to the latter term, meaning that $\tau_1(s)$ does not depend on $n$. 
In other words, it represents the invariant structure of $p_n(x_n)$ inside the peaks. 
Similarly, $1-\tau_1(s)$ is the invariant structure of $q_n(x_n)$.

Congruently, we obtain the following fact about the transition probabilities outside the peaks:
\begin{thm}[\textbf{Outside the peaks}]\label{thm:outside}
	For $s\in\mathbb{R}$ that satisfies $|a|< |s|< 1$ and a sequence $\{x_n\}$ conditioned by Eq.~(\ref{eq:xn1}), $p_n(x_n)$ and $q_n(x_n)$ can be described as
	\begin{align}
		&\lim_{n\rightarrow\infty}p_n(x_n) = \fraction{s-|a|^2 + |b|\sqrt{s^2-|a|^2}}{2s} =:\tau_2(s), \label{eq:pnxn_outside}\\
		&\lim_{n\rightarrow\infty}q_n(x_n) = 1-\tau_2(s).\label{eq:qnxn_outside}
	\end{align}
\end{thm}
Herein, the transition probabilities converge to a function $\tau_2(s)$, without involving time $n$, in contrast to the inside-the-peaks case described above. 

The final statement regards the area {\it around the peaks}, or the boundaries between the regions inside and outside the peaks.  
\begin{thm}[\textbf{Around the peaks}]\label{thm:around}
Let a sequence $\{x_n^{\pm}\}$ satisfy 
\begin{align}\label{eq:xnaround}
	x_{n}^{\pm} = \pm n|a| + d_n^{\pm}, 
\end{align}
where $\{d_n^{{\pm}}\}$ is a sequence that satisfies $d_n^{\pm}= \revise{c} n^{1/3} +o(n^{1/3})$ with any $\revise{c} \geq 0$. Then, it follows that
\begin{align}
    &\lim_{n\to\infty} p_{n}(x_{n}^{\pm}) = \fraction{1\mp |a|}{2} =: \tau_{\circ}^{\pm}, \label{eq:pnx_around_conv}\\
    &\lim_{n\to \infty}q_{n}(x_n^{\pm}) = 1-\tau_\circ^{\pm}.  \label{eq:qnx_around_conv}
\end{align}
\end{thm}

\subsection{Skeleton structure of quantum walks}
The skeleton structure of a {QW} is defined by combining the functions $\tau_1$, $\tau_2$, and $\tau_\circ^{\pm}$ introduced in Theorems~\ref{thm:inside}, \ref{thm:outside}, and \ref{thm:around}:
\begin{defn}[\textbf{Skeleton structure}]\label{thm:skeleton_structure}
	We define the function $\tau: (-1,\,1)\rightarrow [0,\,1]$ as
	\begin{align}\label{eq:skeleton}
		\tau(s) &= \left\{\begin{array}{ll} 
		\tau_1(s) = \fraction{1-s}{2} & (0\leq |s|< |a|)\vrule width 0pt height 9pt depth 12pt \\ \tau_{\circ}^{\pm} = \fraction{1\mp |a|}{2} & (s = \pm |a|)\vrule width 0pt height 9pt depth 12pt \\ \tau_2(s) = \fraction{s-|a|^2 + |b|\sqrt{s^2-|a|^2}}{2s} & (|a|< |s|< 1) \end{array}\right.\hspace{-4.5pt},
	\end{align}
	and call it the \textit{skeleton structure} of {quantum walks}.
\end{defn}
The function $\tau(s)$ defined by Eq. (\ref{eq:skeleton}) is independent of the initial state $\ket{\varphi}$, i.e., the skeleton structure is determined only by the parameters of the coin matrix. 
In addition, on the straight part inside the peaks, the range is determined by the parameters of the coin matrix, but the gradient is always $-1/2$, regardless of the settings of the QWs, as long as the coin matrix is time- and site-homogeneous. That is, the gradient is invariant for all the unitary matrices $C$. 
Moreover, $\tau(s)$ is continuous around the peaks $s=\pm |a|$; that is, 
\begin{align}\label{eq:tau_is_continuous}
\lim_{s\rightarrow \pm |a| \mp 0}\tau_1(s) = \lim_{s\rightarrow \pm |a| \pm 0}\tau_2(s) = \tau_{\circ}^{\pm}.
\end{align}

\begin{figure*}[t] 
\centering\includegraphics[width=\linewidth]{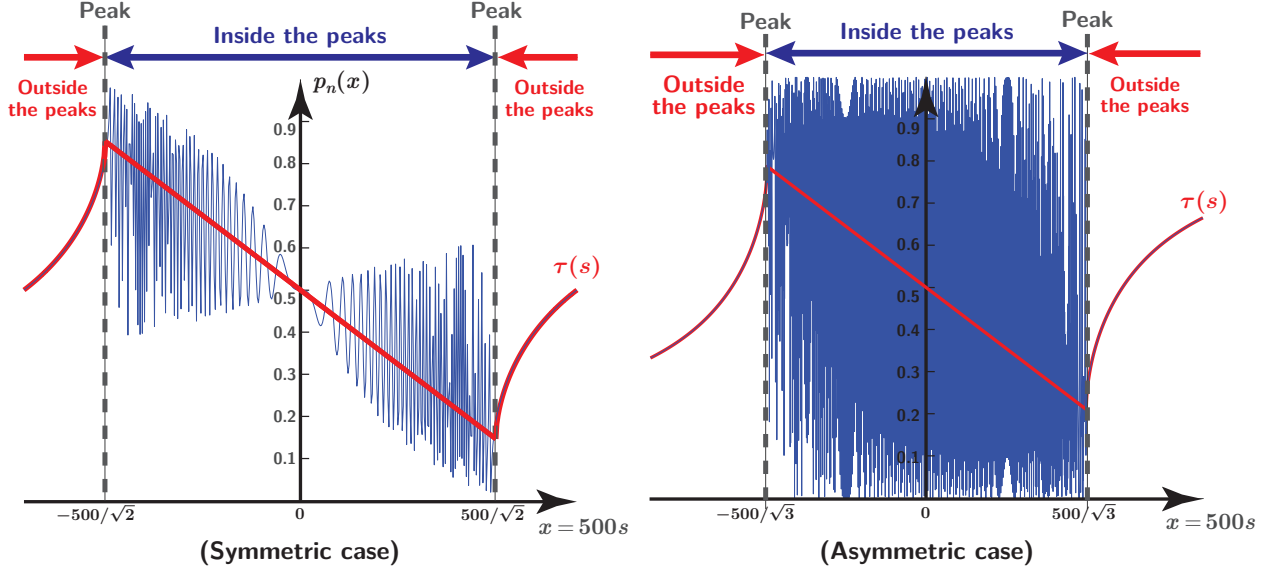}
\caption{The relationship between $p_{500}(x)$ (blue line) and its skeleton structure $\tau(s)$ (red line) in each case.}
\vrule height 0.3mm width 165mm
\label{fig:skeleton}
\end{figure*}

The graph of this skeleton structure $\tau(s)$ fits with that of $p_n(x)$ by stretching the $s$-axis $n$ times. 
As examples, we show the symmetric and asymmetric cases in Figs.~\ref{fig:skeleton}(a) and (b), respectively. 
Inside the peaks, these two cases seemingly exhibit quite different oscillatory behavior. 
However, the skeleton structures of these two cases are the same. 
Outside the peaks, both cases exhibit curves that do not accompany oscillatory behavior. 
This is the visualization of the convergence stated in Theorem~\ref{thm:outside}.

The skeleton structure encapsulates the properties of quantum walks. The function $\tau(s)$ indicates how the walkers tend to move in the left direction, while the right direction corresponds to $1-\tau(s)$. We can capture the characteristic that the walkers tend to the position $x = \pm n|a|$, resulting in the high peaks there. This is more reliable than describing the behavior with the functions $p_n$ and $q_n$; the complicated oscillation that prevents us from quantitative estimation is excluded.


\section{Quantum-skeleton random walk}\label{sec:qsrw}\setcounter{equation}{0}
\begin{figure*}[t]
\begin{tabular}{c}
\begin{minipage}[b]{\linewidth} 
    \centering\includegraphics[width=0.97\linewidth]{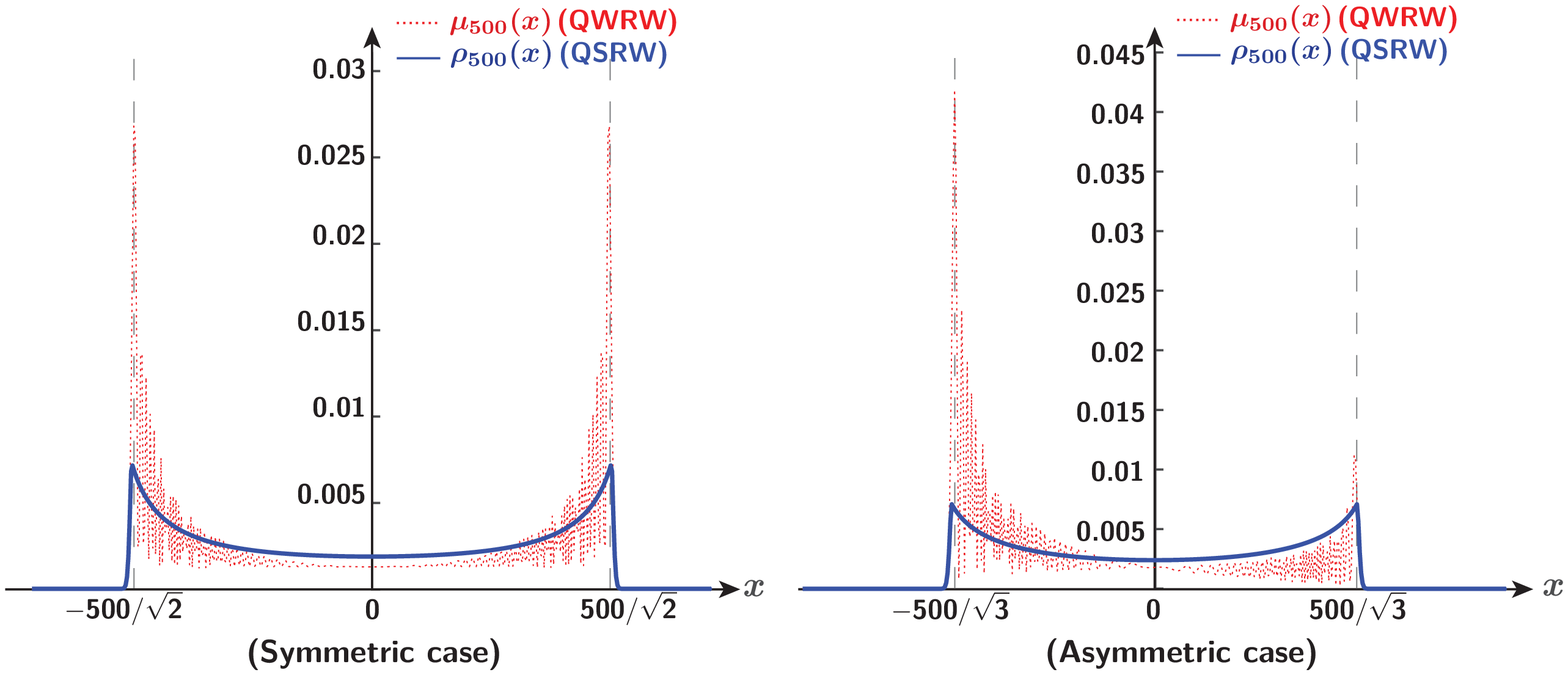}
	\caption{The probability distribution of the QSRW at time $n=500$ ($\rho_{500}(x)$, blue line) and comparison with the that of the corresponding QWRW (${\mu}_{500}(x)$, red line{, dashed}). Each value is plotted only on the even points; on the odd points, the probabilities are $0$.}\label{fig:qsrw}
\vrule height 0.3mm width 165mm\vspace{0.4\baselineskip}
  \end{minipage}\\
  \begin{minipage}[t]{\linewidth} 
   \includegraphics[width=0.97\linewidth]{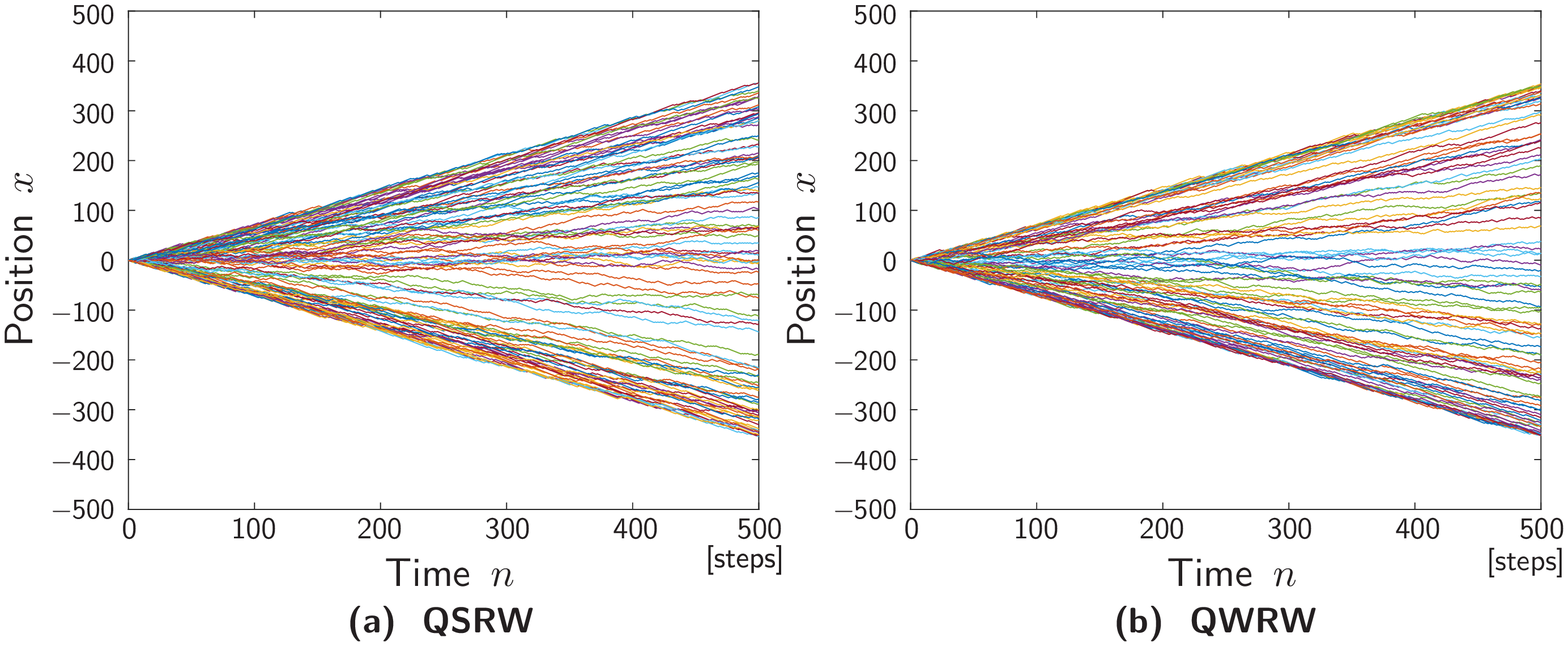}
	\caption{Paths of walkers following (a) QSRW and (b) QWRW (the final time $n=500$, the number of walkers $K=100$). In both cases, paths spread radially, which is caused by linear spreading.}\label{fig:path}
\vrule height 0.3mm width 165mm
  \end{minipage}
 \end{tabular}
\end{figure*}
\begin{figure}[htb] 
\centering\includegraphics[width=0.6\linewidth]{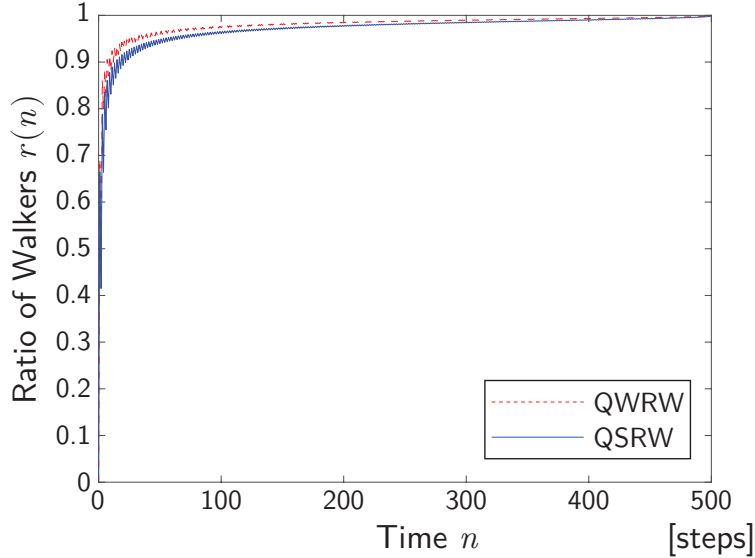}
\caption{Future direction following QSRW (red line; dashed) and QWRW (blue line) \revise{in the symmetric case on Table~\ref{tb:cases}}. The final time is $N=500$.}\label{fig:future}
\vrule height 0.3mm width 165mm
\end{figure}
\begin{figure}[htb] 
\centering\includegraphics[width=0.6\linewidth]{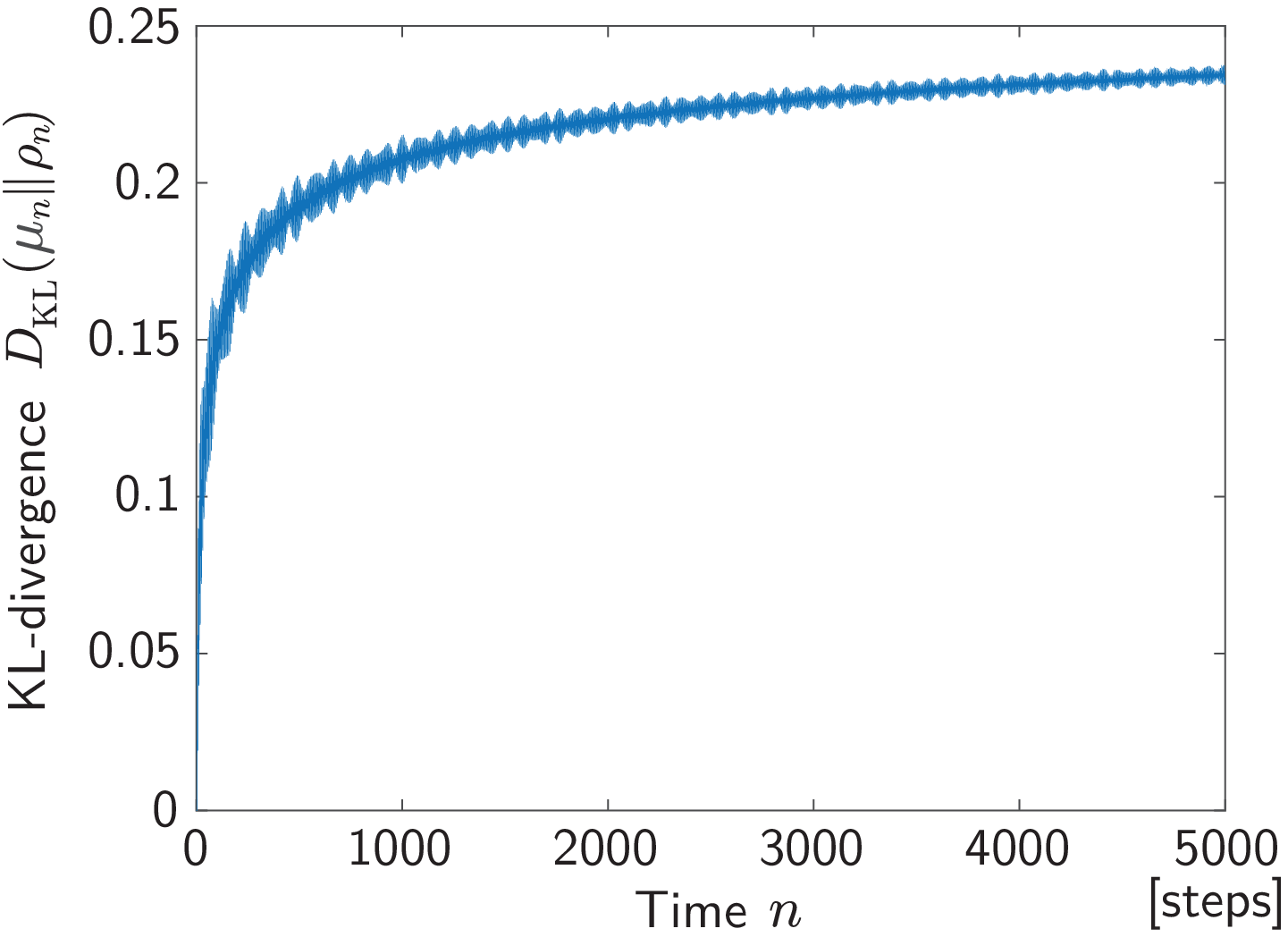}
\caption{{KL-divergence between probability distributions between QWRW and QSRW \revise{in the symmetric case on Table~\ref{tb:cases}}, denoted by $D_{\mathrm{KL}}(\mu_n\|\rho_n)$.}}\label{fig:kl}
\vrule height 0.3mm width 165mm
\end{figure}

To further analyze the properties of the skeleton structure of quantum walks, we now construct a random walk whose transition probabilities are defined \revise{based on only} the skeleton structure of quantum walks{, i.e., without the oscillatory behavior present in QWRWs.}
We call {this new type of} random walk a \textit{quantum-skeleton random walk} (QSRW). 

\revise{The transition probabilities of QSRW is defined by the skeleton structure whose input is the position appropriately scaled by time step $n$. Then it is reasonable to define the distribution of $\tilde{p}_0(0)$ and $\tilde{q}_0(0)$ as $\tau(0)$ by the symmetry; then $p_n(0) = q_n(0) = 1/2$ holds for any $n\in \mathbb{N}_0$.} Precisely, the transition probability of the QSRW to the left (resp. right) side, denoted by $\tilde{p}_n(x)$ (resp. $\tilde{q}_n(x)$), is described as follows: for $n=0$,
\begin{align}
    \tilde{p}_0(0) &= \tau(0) = \fraction{1}{2},\label{eq:qsrwp0}\\
    \tilde{q}_0(0) &= 1-\tilde{p}_0(0) = \fraction{1}{2}.\label{eq:qsrwq0}
\end{align}
For $n\geq 1$,
\begin{align}
	\tilde{p}_n(x) &= \tau(x/n)\nonumber \\
	&= \left\{\begin{array}{ll} \fraction{n-x}{2n} \hspace{0.18\linewidth} (|x| \leq n|a|)\vrule width 0pt height 9pt depth 12pt \\ \fraction{x-n|a|^2 +|b|\sqrt{x^2-n^2|a|^2}}{2x} \\
	\hspace{0.3\linewidth}(n|a| < x \leq n) \end{array}\right., \label{eq:qsrwp}
\end{align}
\begin{align}
	\tilde{q}_n(x) &= 1-\tilde{p}_n(x) = 1- \tau(x/n). \label{eq:qsrwq}
\end{align}

The probability distribution of the QSRW, denoted by $\rho_n(x)$, has the following recurrent formula:
\begin{align}\label{eq:recurrent_rho}
	\rho_{n+1}(x) = \tilde{p}_n(x+1)\rho_n(x+1)  + \tilde{q}_n(x-1)\rho_n(x-1).
\end{align}
In addition, we assume that walkers start from the origin:
\begin{align}\label{eq:init_rho}
	\rho_0(x) = \mathbf{1}_{0}(x).
\end{align}

Figs.~\ref{fig:qsrw} (a) and (b) show comparisons of the probability distribution between the QSRW ($\rho_{500}(x)$, blue line) and the corresponding QWRW (${\mu}_{500}(x)$, red \revise{dashed} line) in the symmetric and asymmetric cases over positions $x\in\mathbb{Z}$ at time $n=500$, respectively. 
In both cases, we observe that the probability distributions $\rho_{500}(x)$ and ${\mu}_{500}(x)$ exhibit linear spreading, with their peaks at the positions $x= \pm 500|a|$.
At the same time, however, the degree of linear spreading in the QSRW is attenuated compared with the QWRW; the maximum probability around the peak in the QSRW exhibits a lower value compared with the QWRW (and thus also with the original QW). 
This implies that the effect of the oscillatory terms of $p_n$ and $q_n$ may contribute to enhancing the linear spreading of the QWs.
{These properties are also observed for a general time step $n$; the peaks of the probability distribution of QSRW are always around $x= \pm n|a|$, and the height is lower than that of the corresponding QWRW.}

Besides, in both cases, $\rho_{500}(x)$ of QSRW is {always} symmetric.
That is, even if the probability distribution of the QWRW is asymmetric (Fig. \ref{fig:qsrw}(b)), the corresponding QSRW results in a symmetric probability distribution. 
\revise{This symmetry} of the QSRW is mathematically derived from the equation $\tilde{p}_n(x) = \tilde{q}_n(-x)$ for all $x$, which is easily obtained from Eqs.~\refeq{eq:qsrwp} and \refeq{eq:qsrwq}. 

\revise{Moreover, the probability distribution of the QWRW, as well as the original QW, shows an oscillatory dependency on the spatial position, especially around the peak. 
By contrast, the probability distribution of the QSRW shows smooth dependency on position; such a property results from the simplified structure of the transition probabilities.
It is one of the interesting observations that this smoothing of the probability distributions occurs in removing the initial state and asymmetry of the peaks.}

Although there are some remarkable differences in probability distributions between the QSRW and QWRW, as described above, we observe similar properties from the viewpoint of \textit{directivity} \cite{TY22a}. \revise{In the following, we investigate it in the symmetric case on Table~\ref{tb:cases}.}

Figs.~\ref{fig:path} (a) and (b) show path trajectories of \revise{100} individual walkers following the QSRW or QWRW in the symmetric case \revise{given in Table~\ref{tb:cases}}, respectively. 
In each case, individual trajectories start from the coordinates $(n,\, x) = (0,\, 0)$; this indicates that the particles are surely on the origin at $n=0$ (${\mu}_0(0) = \rho_0(0) = 1$). 
From these figures, both cases exhibit linear spreading trajectories. 
Certainly, the density of paths of the QSRW at the edges of the range of motion is lower than that of the QWRW, which corresponds to the difference of the height of the peaks between both walks in Fig.~\ref{fig:qsrw} (the symmetric case). However, many more walkers reach positions outside $\pm 300$ than would be expected in the case of a simple random walk. 
The upper and lower bounds of the finite positions of trajectories are about \revise{$\pm 500/\sqrt{2}$}, which corresponds to the peaks of the probability distributions shown in Fig.~\ref{fig:qsrw}.

To examine the similarity between the QSRW and QWRW, Fig.~\ref{fig:future} shows the \textit{future direction} \cite{TY22a} of walkers. 
Here we define a figure-of-merit $r(n)$ of future direction defined by \revise{
\begin{align}\label{eq:rn}
	r(n) = \mathbb{P}(\sgn (S_N \cdot S_n) = 1).
\end{align}
Here, $N\in\mathbb{N}_0$ is the final time instant, and $S_n$ is the position of a walker at time step $n\in \{0,\,1,\,\cdots,\,N\}$ for respective walks. This definition is detailed in Appendix~\ref{app:future}. If a walker chooses the trajectory that satisfies $\sgn (S_N \cdot S_n)=1$ in the definition of $r(n)$ (\ref{eq:rn}), then the walker is on the same side as the final position with respect to the origin at time step $n$. 
On the other hand, if it chooses the trajectory that satisfies $\sgn  (S_N \cdot S_n)= -1$, then the walker is on the opposite side of the final position with respect to the origin at time step $n$. Moreover, if a trajectory gives $\sgn  (S_N \cdot S_n)= 0$, either relation $\sgn (S_n) = 0$ or $\sgn (S_N) = 0$ holds; the latter indicates that a walker goes to neither positive nor negative side in the end. Note that $r(0) = 0$, shown in Fig.~\ref{fig:future}, is derived from the assumption that walkers start from the origin ($S_0 = 0$). 
If a walker maintains the relation ${\rm sgn}(S_N \cdot S_n) =1$, we can say that the walker determines its own evolving direction, and thus the probability of satisfying the equation $\sgn (S_N \cdot S_n) = 1$ can be the benchmark for investigating when walkers effectively determine the directions in which they arrive in the future.}

As represented in the red curve in Fig.~\ref{fig:future}, $r(n)$ of the QWRWs increases dramatically soon after the time evolution begins after $n = 0$. 
That is, the future direction of a walker is highly determined by the position in its early stages. 
This tendency is also clearly observed in QSRWs, as shown by the blue curve in Fig.~\ref{fig:future}; the steep increase of $r(n)$ in the early stages in the QSRW is almost equivalent to that in the QWRW. 
From this analysis, the property of linear spreading of the QSRW is comparable to that of the QWRW. The walkers quickly spread away from the origin while reducing the probability of returning to the origin again.

Finally, we numerically show the similarity of probability distributions between QWRW and QSRW via KL-divergence defined as follows:
\begin{align}\label{eq:kl-div}
	D_{\mathrm{KL}}(\mu_n\|\rho_n) = \sum_{x\in\supp \mu_n}\mu_n(x)\ln\fraction{\mu_n(x)}{\rho_n(x)}
\end{align}
with $\supp\mu_n = \{x\in\mathbb{Z}\,|\,\mu_n(x) >0\}$. 
\revise{The value of $D_{\mathrm{KL}}(\mu_n\|\rho_n)$ is obtained for any $n\in\mathbb{N}$ by iteratively using the recurrence formulae \refeq{eq:rec_mu} and \refeq{eq:recurrent_rho}, as shown in Fig.~\ref{fig:kl}.}
Therein, we can observe that the values in the finite range grow slowly with fluctuations. This indicates that the similarity of the two walks is well maintained, and the effect from neglecting the initial states is weak, at least for the symmetric case studied here.

\section{Proofs of Theorems}\label{sec:proof}\setcounter{equation}{0}
In this section, we use the results obtained by Sunada and Tate \cite{TS12} in proving the theorems described in Sec.~\ref{subsec:statements}. 
The definition used by them is different from our one, and thus we have to make an appropriate transformation from theirs to ours. 

\revise{Let $\varXi_n^{(\delta,\,\sharp)}$ be the weight of all the possible paths from the origin to the position $x$ at time $n$ in the case where $\arg(\det\, C)$ is $\delta \in [-\pi,\,\pi)$, and the type of quantum walk is $\sharp$. Here, we introduce two types of quantum walks: Ambainis \cite{AA01} and Gudder \cite{SG88} types. The former is the one introduced in Sec.~\ref{sec:qw}; the time evolution of the system is described with the unitary matrix
\begin{align*}
U = \bfit{S}\bfit{C},
\end{align*}
as stated in Eq.~\refeq{eq:U}. Here $\bfit{S}$ and $\bfit{C}$ are defined as Eqs. \refeq{eq:shift} and \refeq{eq:coin}, respectively. On the other hand, the latter is the one used in Ref. \cite{TS12}, where the following unitary time evolution operator $U'$ is applied:
\begin{align}\label{eq:UG}
	U' = \bfit{CS}. 
\end{align}
\quad\vspace{-1.3\baselineskip}}

\revise{The other difference between our model and the one in Ref. \cite{TS12} is $\delta$; the former treats the general $\delta$, but in the latter, $\delta$ is set to $0$.}

\revise{In short, to apply the result obtained by Sunada and Tate \cite{TS12}, we have to transform the Ambainis type to the Gudder type, and the general $\delta$ case to the case of $\delta =0$.} Precisely, we use the following lemma whose proof is given in Appendix~\ref{app:tf}:
\begin{lem}\label{lem:tf}
\begin{align}\label{eq:lemma_v1}
\varXi_n^{(\delta,\,\rA)}(x) = e^{i\delta(n+x)/2}\varTheta^{\revise{\dagger}}\varXi_n^{(0,\,\rG)}(x)\varTheta,
\end{align}
where $\varTheta$ is a unitary matrix defined by
\begin{align}\label{eq:Theta}
	\varTheta = \twobytwo{a}{b}{-\overline{b}}{\overline{a}}. 
\end{align}
\end{lem}
In the lemma above, \revise{for a matrix $M$, $M^{\revise{\dagger}}$ represents its conjugate transpose.} 
In the following discussion, we use this lemma to show our results via Sunada and Tate \cite{TS12}. 

For now we assume that $s$ is a rational number on $(-|a|,\,|a|)$ such that $ns\in\mathbb{Z}$ holds. 
However, $s$ can be extended to real numbers by applying the theory of the continued fraction.

The transition probability $p_n(x)$ (see Eq.~\refeq{eq:p}) can be transformed as
\begin{align*}
	p_n(x) &= \fraction{\|P\varPsi_n(x)\|^2}{\mu_n(x)}\\
	&= \fraction{|\braket{\rL|P\varXi_n^\mathrm{(\delta,\,A)}(x)\varphi}|^2}{|\braket{\rL|\varXi_n^\mathrm{(\delta,\,A)}(x)\varphi}|^2 + |\braket{\rR|\varXi_n^\mathrm{(\delta,\,A)}(x)\varphi}|^2}.
\end{align*}
Applying Lemma~\ref{lem:tf}, we have
\begin{align}
	p_n(x) &= \fraction{|\braket{\rL| P\varTheta^{\revise{\dagger}}\varXi_n^\mathrm{(0,\,G)}(x)\varTheta|\varphi}|^2}{|\braket{\rL| \varTheta^{\revise{\dagger}}\varXi_n^\mathrm{(0,\,G)}(x)\varTheta|\varphi}|^2 +|\braket{\rR| \varTheta^{\revise{\dagger}}\varXi_n^\mathrm{(0,\,G)}(x)\varTheta|\varphi}|^2}\nonumber \\
	&= \fraction{|\braket{\rL|\varXi_n^\mathrm{(0,\,G)}(x)|\tilde{\varphi}}|^2}{|\braket{\tilde{\rL}|\varXi_n^\mathrm{(0,\,G)}(x)|\tilde{\varphi}}|^2 +|\braket{\tilde{\rR}|\varXi_n^\mathrm{(0,\,G)}(x)|\tilde{\varphi}}|^2}.\label{eq:pnx}
\end{align}
It should be noted that $P\varTheta\ket{\rL} = \ket{\rL}$ holds. Moreover, we set $\ket{\tilde{\,\cdot\,}} = \varTheta\ket{\,\cdot\,}$, and then because of the unitarity of $\varTheta$, $\|\tilde{\varphi}\|=1$ holds. 

Here, we introduce the functions of vectors $\ket{u(z)}$ and $\ket{v(z)}$ with $z \in S^1 := \{z\in\mathbb{C}\,|\,|z|=1\}$:
	\begin{align}
		&\ket{u(z)} = \fraction{1}{N(z)}\twobyone{abz/|a|}{\revise{h(z)} -|a|\overline{z}},\label{eq:u}\\
		&\ket{v(z)} = \fraction{1}{N(z)}\twobyone{-\revise{\overline{h(z)}} +|a|z}{\overline{abz}/|a|},\label{eq:v}
	\end{align}
where $N(z)$ is a normalizing function, and
\begin{align}\label{eq:h}
    h(z) = |a|\cos(\arg z) + i\sqrt{1-|a|^2\cos^2(\arg z)}.
\end{align}
Note that, assuming $\ket{u(z)} = [u_{\rL}(z)\,\,\,u_{\rR}(z)]^\trp$, it follows that
\begin{align}\label{eq:v_by_u}
	\ket{v(z)} = \twobyone{-\overline{u_{\rR}(z)}}{\overline{u_{\rL}(z)}}.
\end{align}

\subsection{Inside the peaks (Theorem~\ref{thm:inside})}\label{subsec:inside}
Let $k(s)$ and $\omega(s)$ be $\mathbb{R}$-valued functions defined as follows:
\begin{align}
	&k(s) = \arcsin\fraction{|b|s}{|a|\sqrt{1-s^2}},\label{eq:k}\\
	&\omega(s) = \arccos(|a|\cos k(s)). \label{eq:omega}
\end{align}
With the functions above, we use the following proposition:
\begin{prp}[Proposition 2.2 in \cite{TS12}]\label{prp:inside}
Let $x \in\mathbb{Z}$ and $s_n \in (-|a|,\,|a|)$ satisfy $s_n = {x}/{n}$.
Then, for $\ket{\psi} \in \mathbb{C}^2$ and $\ket{\chi}\in\mathbb{C}^2$ with $\|\psi\| = \|\chi\| = 1$, we have
\begin{align*}
	&\braket{\psi| \varXi_n^{(0,\,\rG)}(x) |\chi} \\
	=& \left( 1+(-1)^{n+x}\right) e^{-x\arg a}\sqrt{\fraction{|b|}{2\pi n(1-s_n^2)\sqrt{|a|^2 -s_n^2}}}\cdot\left( e^{i(n\theta(s_n)+\pi/4)}f_\chi^\psi(s_n) + e^{-i(n\theta(s_n)+\pi/4)}g_\chi^\psi(s_n) + O(1/n)\right),
\end{align*} 
where
\begin{align}
	\theta(s) &= \omega(s) - sk(s), \label{eq:theta}\\
	 f_\chi^\psi(s) &= \braket{\psi| u(e^{ik(s)})} \braket{u(e^{ik(s)}) | \chi},\label{eq:fpsi}\\
	g_\chi^\psi(s) &= \braket{\psi| v(e^{ik(s)})} \braket{v(e^{ik(s)}) | \chi}.\label{eq:gpsi}
\end{align}
\end{prp}

Applying the result of Proposition~\ref{prp:inside} to Eq.~(\ref{eq:pnx}), we obtain
\begin{align}\label{eq:pnx_fg}
	p_n(x) = \fraction{G^{\revise{\rL}}(s_n) + \tilde{\xi}_n(s_n) + O(1/n)}{G^{\revise{\tilde{\rL}}}(s_n) + G^{\revise{\tilde{\rR}}}(s_n) + \tilde{\eta}_n(s_n) + O(1/n)}, 
\end{align}
where
\begin{align}
    &G^{\revise{\chi}}(s) = |f_{\revise{\tilde{\varphi}}}^{\revise{\chi}}(s)|^2 + |g_{\revise{\tilde{\varphi}}}^{\revise{\chi}}(s)|^2, \label{eq:G}\\
	&\tilde{\xi}_n(s)  = \revise{-2\im \left( e^{2in\theta(s)}f_{\revise{\tilde{\varphi}}}^{\rL}(s)\overline{g_{\revise{\tilde{\varphi}}}^{\rL}(s)}\right)}, \label{eq:xitilde}\\ 
	&\tilde{\eta}_n(s) = \revise{-2 \im \left(e^{2in\theta(s)}(f_{\revise{\tilde{\varphi}}}^{\tilde{\rL}}(s)\overline{g_{\revise{\tilde{\varphi}}}^{\tilde{\rL}}(s)} +f_{\revise{\tilde{\varphi}}}^{\tilde{\rR}}(s)\overline{g_{\revise{\tilde{\varphi}}}^{\tilde{\rR}}(s)})\right)}. \label{eq:etatilde} 
\end{align}
Here, by simple calculation, we obtain
\begin{align*}
	&\revise{G^\rL (s_n)} = \tau_1(s_n)\lambda(s_n),\\
	&\revise{G^{\tilde{\rL}} (s_n) + G^{\tilde{\rR}}(s_n)}  = \lambda(s_n),
\end{align*}
where $\tau_1(s)$ and $\lambda(s)$ are given as follows:
\begin{align}
	\tau_1(s)&=\fraction{1-s}{2}, \label{eq:taus}\\
	\lambda(s) &= 1-\left(|\alpha|^2-|\beta|^2 +\fraction{2\Re(a\alpha\overline{b\beta})}{|a|^2}\right)s, \label{eq:lambdas}
\end{align}
\revise{assuming that $\ket{\varphi} = [\alpha\,\,\,\beta]^\trp$. Note that $|\alpha|^2 + |\beta|^2 = 1$.}

Since $\lambda(s) \not =0$ for $s\in (-|a|,\,|a|)$, Eq.~(\ref{eq:pnx_fg}) can be transformed to
\begin{align}\label{eq:pnx_fin}
	p_n(x) = \fraction{\tau_1(s_n) + \tilde{\xi}_n(s_n)/\lambda(s_n)}{1+ \tilde{\eta}_n(s_n)/\lambda(s_n)} = \fraction{\tau_1(s_n) + \xi_n(s_n)}{1 +\eta_n(s_n)},
\end{align}
where $\xi_n(s) := \tilde{\xi}_n(s)/\lambda(s)$ and $\eta_n(s) := \tilde{\eta}_n(s)/\lambda(s)$.

Here, we can replace the pair $(x,\,s_n)$ with another one $(x_n,\,s) \in\mathbb{Z}\times (-|a|,\,|a|)$ conditioned by (\ref{eq:xn1}), see Eqs.~(1.13) and (1.14) in Sunada and Tate \cite{TS12}.

We show that the weak convergence of $\xi_n(s)$ and $\eta_n(s)$. Since $\lambda(s)$ is independent of $n$, it is sufficient that we prove
\begin{align}\label{eq:wlim_of_xi_eta}
	\wlim_{n\to\infty}\tilde{\xi}_n(s) = \wlim_{n\to\infty}\tilde{\eta}_n(s) = 0.
\end{align}
Actually, \revise{for $\tilde{\xi}_n(s)$, there exist continuous functions $F_1^{(\xi)}$ and $F_2^{(\xi)}$ such that
\begin{align}\label{eq:osc_func_xi}
	\tilde{\xi}_n(s) = F_1^{(\xi)}(s)\sin(2n\theta(s)) + F_2^{(\xi)}(s)\cos(2n\theta(s)).
\end{align}
Equally, $\tilde{\eta}_n(s)$, there also exist 
continuous functions $F_1^{(\eta)}$ and $F_2^{(\eta)}$ such that
\begin{align}\label{eq:osc_func_eta}
	\tilde{\eta}_n(s) = F_1^{(\eta)}(s)\sin(2n\theta(s)) + F_2^{(\eta)}(s)\cos(2n\theta(s)).
\end{align}}
This can be claimed from the fact that the trigonometric functions of $k(s)$ and $\omega(s)$ are continuous, and \revise{$F_1^{(\natural)}(s)$ and $F_2^{(\natural)}(s)$} are obviously written with them \revise{($\natural = \xi,\ \eta$)}.

Here, we introduce the following proposition, whose proof is shown in Appendix~\ref{app:rl}:

\begin{prp}[Extension of Riemann-Lebesgue lemma]\label{prp:rl}\quad
	For any $y\in[-|a|,\,|a|]$ and a continuous function $F:[-|a|,\,|a|]\to \mathbb{C}$,
	\begin{align}
		\lim_{n\rightarrow\infty}\int_{-|a|}^{y}F(s)\sin(2n\theta(s)) \rd s = \lim_{n\rightarrow\infty}\int_{-|a|}^{y}F(s)\cos(2n\theta(s)) \rd s = 0.\nonumber
	\end{align}
\end{prp}
Applying this to Eqs.~\refeq{eq:osc_func_xi} and \refeq{eq:osc_func_eta}, we have
\begin{align}\label{eq:apply_v3}\begin{split}
	\lim_{n\rightarrow\infty}\int_{-|a|}^{y}(F_1^{\revise{(\natural)}}(s)\sin(2n\theta(s)) + F_2^{\revise{(\natural)}}(s)\cos(2n\theta(s)))\rd s= 0.
\end{split}\end{align}
By the definition of weak convergence, this is equivalent to
\begin{align}\label{eq:weak_limit}
	\wlim_{n\to\infty} (F_1^{\revise{(\natural)}}(s)\sin(2n\theta(s)) +F_2^{\revise{(\natural)}}(s)\cos(2n\theta(s))) = 0.
\end{align}
Therefore, we conclude that \revise{Eq.~\refeq{eq:wlim_of_xi_eta} holds;} that is,
\begin{align}\label{eq:target}
\wlim_{n\to\infty}{\xi}_n(s) = \wlim_{n\to\infty}{\eta}_n(s) = 0.
\end{align}

The right transition probability, denoted by $q_n(x)$, is described as
\begin{align}\label{eq:qnx_xi_eta}
	q_n(x) &= 1- p_n(x) = \fraction{1-\tau_1(s_n) +\eta_n(s_n)-\xi_n(s_n)}{1+\eta_n(s_n)},
\end{align}
where we applied Eq.~(\ref{eq:pnx_fin}). Defining $\zeta_n(s_n) := \eta_n(s_n)-\xi_n(s_n)$, we obtain
\begin{align}\label{eq:wlim_of_zeta}
	\wlim_{n\to\infty} \zeta_n(s) = 0
\end{align}
by the linearity of integration and the limit. 

From the above, we obtain the desired result.

\subsection{Outside the peaks (Theorem~\ref{thm:outside})}
We use the following proposition:
\begin{prp}[Proposition 4.1 in \cite{TS12}]\label{prp:s41}
	Let $s$ satisfy $|a|<|s|<1$, and a sequence $\{x_n\}$ be conditioned by (\ref{eq:xn1}). For any $\ket{\psi}\in\mathbb{C}^2$ and $\ket{\chi} \in\mathbb{C}^2$ with $\|\psi\| = \|\chi\|=1$,
	\begin{align*}
		&\braket{\psi|\varXi_n^{\rm (0,\,G)}(x_n)|\chi} \nonumber \\
		&=\{ 1+(-1)^{n+x_n}\}\fraction{e^{-x_n\arg a}e^{-nH_{\rm Q}(s)/2}}{\sqrt{2\pi n}}\sqrt{\fraction{|b|}{(1-s^2)\sqrt{s^2-|a|^2}}}\times e^{\pi i(n-|x_n|)/2}(F_{\psi}(s)+O(1/n)),
	\end{align*}
	where $H_{\mathrm{Q}}(s)$ is a convex positive-valued function given by
	\begin{align*}
		H_{\mathrm{Q}}(s) =&2|s|\ln (|bs|+\sqrt{s^2-|a|^2})-2\ln (|b|+\sqrt{s^2-|a|^2})+(1-|s|)\ln (1-s^2) -2|s|\ln |a|,
	\end{align*}
	and using
	\begin{align*}
		r(s) = \fraction{|b|s + \sqrt{s^2-|a|^2}}{|a|\sqrt{1-s^2}}
    \end{align*}
    and
    \begin{align*}
        D(s) = \fraction{|b|+\sqrt{s^2-|a|^2}}{\sqrt{1-s^2}},
	\end{align*}
	$F_\psi(s)$ is defined as
	\begin{align*}
		F_{\psi}(s) = \phi(s) \braket{\psi|w(s)}
	\end{align*}
	with
	\begin{align*}
		&\phi(s) = \fraction{i|a|\left(|a| r(s)^{-1}-D(s)^{-1}\right) \chi_{\rL}-ir(s) a b \chi_{\rR}}{r(s) a b\left(D(s)+D(s)^{-1}\right)},\\
		&\ket{w(s)} = i\twobyone{r(s)ab/|a|}{D(s)+|a|r(s)^{-1}}.
	\end{align*}
\end{prp}

First of all, similarly to the case around the peaks, $p_n(ns)$ can be transformed to
\begin{align*}
		p_n(x_n) = \fraction{|\braket{{\rL}|\varXi_n^{\rm (0,\,\rG)}(x)|\tilde{\varphi}}|^2}{|\braket{\tilde{\rL}|\varXi_n^{\rm (0,\,\rG)}(x)|\tilde{\varphi}}|^2 + |\braket{\tilde{\rR}|\varXi_n^{\rm (0,\,\rG)}(x)|\tilde{\varphi}}|^2}.		
\end{align*}
By Proposition~\ref{prp:s41},
\begin{align*}
		p_n(x_n) =& \fraction{|F_{\rL}(s)|^2 + O(n^{-2})}{|F_{\tilde{\rL}}(s)|^2+ |F_{\tilde{\rR}}(s)|^2 +O(n^{-2})}\nonumber\\
		=& \fraction{1}{|F_{\tilde{\rL}}(s)|^2+ |F_{\tilde{\rR}}(s)|^2} (|F_{\rL}(s)|^2 + O(n^{-2}))\left(1 + \fraction{O(n^{-2})}{|F_{\tilde{\rL}}(s)|^2+ |F_{\tilde{\rR}}(s)|^2}\right)^{-1}\nonumber\\
		=& \fraction{|F_{\rL}(s)|^2}{|F_{\tilde{\rL}}(s)|^2+ |F_{\tilde{\rR}}(s)|^2} +O(n^{-2}).
\end{align*}
Therefore,
\begin{align*}
	\lim_{n\rightarrow\infty}p_n(x_n) &= \fraction{|F_{\rL}(s)|^2}{|F_{\tilde{\rL}}(s)|^2+ |F_{\tilde{\rR}}(s)|^2}= \fraction{\phi(s)\braket{\rL|w(s)}|^2}{|\phi(s)\braket{\tilde{\rL}|w(s)}|^2 + |\phi(s)\braket{\tilde{\rR}|w(s)}|^2}\nonumber\\
		&= \fraction{|\braket{\rL|w(s)}|^2}{\|w(s)\|^2} = \fraction{s-|a|^2 + |b|\sqrt{s^2-|a|^2}}{2s}.
\end{align*}
We can also confirm convergence regarding $q_{n}(x)$ easily from the property of the limit.

\subsection{Around the peaks (Theorem~\ref{thm:around})}
We use the following proposition:
\begin{prp}[Proposition 3.1 in \cite{TS12}]\label{prp:s31}
	Let a sequence $\{x_n^{\pm}\}$ satisfy
	\begin{align*}
		x_n^{\pm} = \pm n|a| + d_n,
	\end{align*}
	where a sequence $\{d_n\}$ satisfies $d_n = O(n^{1/3})$. For any $\ket{\psi}\in\mathbb{C}^2$ and $\ket{\chi}\in\mathbb{C}^2$ with $\|\psi\| = \|\chi\|=1$,
	\begin{align*}
		\braket{\psi|\varXi_n^{(0,\,\mathrm{G})}(x_n^{\pm})|\chi}= (1+(-1)^{n+x_n^{\pm}})e^{-ix_n^{\pm}\arg(a)}e^{i\pi (n\mp x_n^{\pm})/2}\times f_\chi^\psi(\pm i) \kappa n^{-1/3} \mathrm{Ai}(\pm \kappa n^{-1/3}d_n) + O(n^{-2/3}).
	\end{align*}
Here,
\begin{align*}
	\kappa = \left(\fraction{2}{|a||b|^2} \right)^{1/3},\quad \tau_\circ^{\pm} = \fraction{1\mp |a|}{2},
\end{align*}
and $\mathrm{Ai}(z)$ is the Airy function defined as
\begin{align*}
	\mathrm{Ai}(z) &= \fraction{1}{2\pi} \int_{-\infty}^{\infty} e^{it^3/3 +itz}\,\rd t= \fraction{1}{\pi}\int_{0}^{\infty} \cos\left(\fraction{t^3}{3} + tz \right) \rd t.
\end{align*}
	The function $f_\chi^\psi(z)$ is given in Proposition~\ref{prp:inside}.
\end{prp}
Here we choose $\{d_n^{\pm}\}$ defined by \refeq{eq:xnaround} as $\{d_n\}$ in the proposition above. Applying the result of Proposition~\ref{prp:inside} to Eq.~(\ref{eq:pnx}), we obtain
\begin{align}
	p_n(x_n^{\pm}) =\,& \fraction{|f_{\varphi}^{\rL}(\pm i) \mathrm{Ai}(\pm \kappa n^{-1/3}d_n^{\pm})|^2 +O(n^{-2/3})}{(|f_{\varphi}^{\tilde{\rL}}(\pm i)|^2 +|f_{\varphi}^{\tilde{\rR}}(\pm i)|^2) |\mathrm{Ai}(\pm \kappa n^{-1/3}d_n^{\pm})|^2 +O(n^{-2/3})}\nonumber\\
	=\,& \fraction{|\braket{\rL|u(\pm i)}|^2 |\mathrm{Ai}(\pm \kappa n^{-1/3}d_n^{\pm})|^2 +O(n^{-2/3})}{\|u(\pm i)\|^2 |\mathrm{Ai}(\pm \kappa n^{-1/3}d_n^{\pm})|^2 +O(n^{-2/3})}\nonumber\\
	=\,& \fraction{(1+|a|)(1-|a|) |\mathrm{Ai}(\pm \kappa n^{-1/3}d_n^{\pm})|^2 +O(n^{-2/3})}{2(1\pm |a|) |\mathrm{Ai}(\pm \kappa n^{-1/3}d_n^{\pm})|^2 +O(n^{-2/3})}\nonumber\\
	=\,& \tau_{\circ}^{\pm}\times \fraction{|\mathrm{Ai}(\pm \kappa n^{-1/3}d_n^{\pm})|^2 +O(n^{-2/3})}{|\mathrm{Ai}(\pm \kappa n^{-1/3}d_n^{\pm})|^2 +O(n^{-2/3})}, \label{eq:pnx_around}
\end{align}
where $\tau_\circ^{\pm} = (1\mp |a|)/2$.

Here, the Airy function $\mathrm{Ai}(z)$ takes positive values for all $z>0$, and $\kappa$ is surely positive under the condition $0<|a|<1$. These result in the fact that the Airy function in \refeq{eq:pnx_around} is positive as $n\to\infty$. Therefore, we can cancel  $|\mathrm{Ai}(\pm \kappa n^{1/3}d_{n})|^2$ from Eq.~(\ref{eq:pnx_around}):
\begin{align}\label{eq:pnx_fraction_order}
	p_{n}(x_{n}^{\pm}) = \tau_\circ^\pm\times \fraction{1 + O(n^{-2/3})}{1 + O(n^{-2/3})}.
\end{align}
Moreover, we can rewrite the equation above as
\begin{align}\label{eq:pnx_order}
	p_{n}(x_{n}^{\pm}) = \tau_\circ^\pm + O(n^{-2/3}),
\end{align}
and thus the convergence (\ref{eq:pnx_around_conv}) holds. 
Under this condition, we can also confirm convergence regarding $q_{n}(x_{n}^{\pm})$ easily by the property of the limit.

\section{Conclusion and discussion}\label{sec:summary}\setcounter{equation}{0}
In this paper, we have {newly introduced an underlying structure of quantum walks, which we call the skeleton structure,} in the case of a homogeneous coin matrix.
{The skeleton structure is based on a quantity of quantum walks, which can be interpreted as the transition probabilities of random walks that replicate the probability distribution of quantum walks (quantum-walk-replicating random walk; QWRW).}
It is noteworthy that the skeleton structure is {independent of the initial state and, partially, of the coin matrix}, whereas the base functions $p_n$ and $q_n$ differ significantly depending {on the conditions of quantum walks. That enhances observing the mechanism of quantum walks through the directivity of QWRWs.}
Moreover, we have defined a new time- and site-dependent random walk, whose transition probabilities are directly derived by the skeleton structure, which we named quantum-skeleton random walk (QSRW).
While the QSRW does not perfectly reproduce the probability distribution of the original quantum walk, we have demonstrated that it exhibits linear spreading comparable to the genuine quantum walk in terms of directivity.

There remain a lot of exciting topics related to $p_n$ and $q_n$, or transition probabilities of QWRW. 
{For example}, the oscillation of the transition probabilities can be considered further. 
\revise{Functions $p_n(x)$ and $q_n(x)$ have both spatial and temporal oscillations, which make them complex. Analyzing this structure is, however, important to investigate how the initial state---the element excluded by introducing the skeleton structure---contributes to deriving the probability distribution of quantum walks in terms of QWRW. We have conducted preliminary analyses on the oscillatory behavior, although there are unexplored aspects including the initial-state dependencies. The analyses are given in Appendix~\ref{app:oscillation}.}

\revise{Besides, considering to construct skeleton structures of quantum walks for the models including localization such as one-defect or three-state ones will be necessary. Especially, for a one-defect case,} characteristic transition probabilities have been observed, as reported in \cite{TY22a}. 
Even though for now we have \revise{ignored the case where both linear spreading and localization occur}, more general skeleton structure may yet exist.
The understanding of such universal underlying structures is of significant interest for further narrowing down, how the special properties of quantum walks arise. 

Further studies focusing on QSRWs may also lead to important results. 
For example, \revise{can we obtain the closed formula or limit distribution of $\rho_n$? If we can address this, then we can describe KL-divergence $D_{\mathrm{KL}}(\mu_n\|\rho_n)$ with a closed formula or obtain the limit of it. That will make it easier to investigate the velocity of convergence of $D_{\mathrm{KL}}(\mu_n\|\rho_n)$, which is important to realize quantitative estimation on a deeper level. Besides, we may explain the cause of fluctuations of $D_{\mathrm{KL}}(\mu_n\|\rho_n)$ via the closed formula of $\rho_n$. It is  conjectured that these fluctuations are derived from the fact that the probability distribution of QWRW $\mu_n$ have fluctuations unlike that of QSRW, but quantitative investigation is needed to ensure that.}
One remaining difficulty in analyzing a QSRW lies in how to treat the transition probability outside the peaks.
Actually, inside the peaks, it is known that the transition probability makes walkers spread \revise{linearly}, like the case of elephant random walks \cite{GS04}. However, it is not trivial that this spreading is also observed in a QSRW; outside the peaks, a similar discussion is difficult because $\tilde{p}_n$ in this region depends on the position $x$ in a more complicated fashion. Further considerations on how to analyze $\tilde{p}_n$ and $\tilde{q}_n$ outside $x=\pm n|a|$ could likely yield more rigorous results.

\def\theequation {\theapp.\arabic{equation}}
\def\thethm{\theapp.\arabic{thm}}
\def\thesubsection{\theapp.\arabic{subsection}}

\appndx{Proof of Lemma~\ref{lem:tf}}{app:tf}\setcounter{equation}{0}
Here we give the proof of the transformation formula presented in Section~\ref{sec:proof}. As mentioned in Section~\ref{sec:qw}, there are several ways of defining a unitary time evolution operator or decomposing the coin matrix $C$.  One way presented in Section~\ref{sec:qw} is called the \textit{Ambainis type} \cite{AA01}; the unitary time evolution operator $U$ is defined by Eq.~\refeq{eq:U}, and then the coin matrix $C$ is decomposed as in Eqs.~\refeq{eq:decompp} and \refeq{eq:decompq}. On the other hand, the way used in Ref. \cite{TS12} is called the \textit{Gudder type}, which treats the following unitary time evolution operator:
\begin{align*}
	U' = \bfit{CS},
\end{align*}
\revise{as stated in Eq.~\refeq{eq:UG}. Here,} $\bfit{C}$ and $\bfit{S}$ are the matrices defined by Eqs.~\refeq{eq:coin} and \refeq{eq:shift}, respectively. Then, we can rewrite $U'$ as
\begin{align}\label{eq:UPQ_G}
	U' = S^{-1}\otimes P' + S\otimes Q',
\end{align}
where
\begin{align}
	P' &= C\ketbra{\rL}{\rL} = \twobytwo{a}{0}{-e^{i\delta}\overline{b}}{0},\label{eq:decompGp}\\
	Q' &= C\ketbra{\rR}{\rR} = \twobytwo{0}{b}{0}{e^{i\delta}\overline{a}}. \label{eq:decompGq}
\end{align}
This indicates that the coin matrix $C$ in Eq.~\refeq{eq:coin} is decomposed as in Eqs.~\refeq{eq:decompGp} and \refeq{eq:decompGq}, which is different from the Ambainis type. In particular, the order of the coin matrix and projector operator is switched. Moreover, in Ref. \cite{TS12}, $\delta$ is set to $0$. In short, to apply the result obtained by Sunada and Tate \cite{TS12}, we have to transform the Ambainis type to the Gudder type, and the general $\delta$ case to the case of $\delta =0$.

From now on, we use the following notations:
\begin{enumerate}[$\bullet$]
	\setlength{\leftskip}{20pt}
	\item $C(\delta) = \twobytwo{a}{b}{-e^{i\delta}\overline{b}}{e^{i\delta}\overline{a}}$: the coin matrix. It should be noted that $\arg(\det C(\delta)) = \delta$ and $C(0) = \varTheta$, given by \refeq{eq:Theta}.
	\item $\bfit{C}(\delta) := \sum_{x\in\mathbb{Z}}\ketbra{x}{x}\otimes C(\delta)$.
	\item $U^{(\delta,\,\sharp)}$: the unitary time evolution operator in the type $\sharp$ with the coin matrix $C(\delta)$: $U^{(\delta,\,\rA)} $ and $U^{(\delta,\,\rG)}$ are $U$ and $U'$ given by  Eqs.~\refeq{eq:U} and \refeq{eq:UG}, respectively.
	\item $P^{(\delta,\,\sharp)},\ Q^{(\delta,\,\sharp)}$: the decomposition matrices of $U^{(\delta,\,\sharp)}$ in the type $\sharp$. They correspond to the left and right transitions, respectively. $(P^{(\delta,\,\rA)},\, Q^{(\delta,\,\rA)})$ and $(P^{(\delta,\,\rG)},\, Q^{(\delta,\,\rG)})$ are $(P,\, Q)$ and $(P',\,Q')$ given by \refeq{eq:decompp}, \refeq{eq:decompq}, \refeq{eq:decompGp}, and \refeq{eq:decompGq}, respectively.
	\item $\varXi_n^{(\delta,\,\sharp)}$: the weight of all the possible paths from the origin to position $x$ at time $n$ in the type $\sharp$ with the coin matrix is $C(\delta)$: $\varXi_n^{(\delta,\,\rA)}$ is $\varXi_n$ given by \refeq{eq:varXi}.
\end{enumerate}

First, we show the transformation from the Ambainis type to the Gudder type. From Eq.~\refeq{eq:UG}, the matrix $U^{(\delta,\,\rG)}$ is written as
\begin{align}
	U^{(\delta,\,\rG)} &= \bfit{C}(\delta)\bfit{S}. \label{eq:UGd}
\end{align}
Here, from Eqs.~\refeq{eq:U} and \refeq{eq:UGd},
\begin{align}\label{eq:AG_U}
	U^{(\delta,\,\rA)} = \bfit{C}(\delta)^{\revise{\dagger}}U^{(\delta,\,\rG)}\bfit{C}(\delta)
\end{align}
holds. Hence, it follows that
\begin{align}
	\left(U^{(\delta,\,\rA)}\right)^n = \bfit{C}(\delta)^{\revise{\dagger}}\left(U^{(\delta,\,\rG)}\right)^n \bfit{C}(\delta).
\end{align}
On the other hand, the relation
\begin{align}
	\left(U^{(\delta,\,\sharp)}\right)^n = \sum_{x\in\mathbb{Z}} (S^x\otimes \varXi_n^{(\delta,\,\sharp)}(x))
\end{align}
follows. By applying Eq.~\refeq{eq:AG_U} to both sides of the equation above, we have
\begin{align}
	\varXi_n^{(\delta,\,\rA)}(x) = C(\delta)^{\revise{\dagger}}\varXi_n^{(\delta,\,\rG)}(x)C(\delta), \label{eq:trans1}
\end{align}
which represents transformation from the Ambainis type to the Gudder type.

Next, we show the transformation from the general $\delta$ case to the case of $\delta=0$. We introduce the Fourier transform of $\varXi^{(\delta,\,\sharp)}_n(x)$ as
\begin{align}
	\hat{\varXi}^{(\delta,\,\sharp)}_n(k) = \sum_{x\in\mathbb{Z}}e^{ikx}\varXi^{(\delta,\,\sharp)}_n(x).
\end{align}
Then, we also have
\begin{align}
	\varXi^{(\delta,\,\sharp)}_n(x) = \int_{-\pi}^{\pi}e^{-ikx}\hat{\varXi}^{(\delta,\,\sharp)}_n(k)\fraction{\rd k}{2\pi} \label{eq:hatXiev}
\end{align}
and call it the Fourier inverse transform. From Eq.~\refeq{eq:tevXi},
\begin{align}
	\hat{\varXi}^{(\delta,\,\sharp)}_n(k) = \hat{U}^{(\delta,\,\sharp)}(k)\,\hat{\varXi}^{(\delta,\,\sharp)}_{n-1}(k) \label{eq:hatvarXi}
\end{align}
with
\begin{align}
	\hat{U}^{(\delta,\,\sharp)}(k) = e^{-ik}P^{(\delta,\,\sharp)} +e^{ik}Q^{(\delta,\,\sharp)}. \label{eq:hatU}
\end{align}
Assigning this to Eqs.~\refeq{eq:hatXiev} and \refeq{eq:hatvarXi} repeatedly, we have
\begin{align}
	\varXi^{(\delta,\,\sharp)}_n(x) &= \int_{-\pi}^{\pi}e^{-ikx}(\hat{U}^{(\delta,\,\sharp)}(k))^n\hat{\varXi}^{(\delta,\,\sharp)}_0(k)\fraction{\rd k}{2\pi} = \int_{-\pi}^{\pi}e^{-ikx}(\hat{U}^{(\delta,\,\sharp)}(k))^n\fraction{\rd k}{2\pi}. \label{eq:inverseXiU}
\end{align}
Note that $\hat{\varXi}^{(\delta,\,\sharp)}_0(k) = I_2$, which is the identity matrix of order 2, since the quantum walk starts from the origin.

On the other hand, by simple calculation, Eq.~\refeq{eq:hatU} can be transformed to
\begin{align}
	\hat{U}^{(\delta,\,\sharp)}(k) = e^{i\delta/2}\,\hat{U}^{(0,\,\sharp)}\left(k +\fraction{\delta}{2}\right).
\end{align}
Applying this to Eq.~\refeq{eq:inverseXiU}, we obtain
\begin{align}
	\varXi^{(\delta,\,\sharp)}_n(x) &= \int_{-\pi}^{\pi} e^{-ikx}\,e^{in\delta/2}\,\hat{U}^{(0,\,\sharp)}\left(k +\fraction{\delta}{2}\right)^n \fraction{\rd k}{2 \pi}= e^{i\delta(n+x)/2}\, \varXi^{(0,\,\sharp)}_n(x), \label{eq:trans2}
\end{align}
which implies transformation from the case with $C(\delta)$ to the case with $C(0)$.

Combining Eqs.~\refeq{eq:trans1} and \refeq{eq:trans2}, we obtain the desired result.

\appndx{More Rigorous Definition of Future Direction}{app:future}\setcounter{equation}{0}
Let $\Omega_N \subset \{\pm 1\}^N$ be the set of possible path trajectories of walks until time step $N$. That is, an element $\bfit{\omega} := (\omega_0,\,\omega_1,\,\cdots,\,\omega_{N-1})\in \Omega_N$ represents which direction is chosen at respective time step; $\omega_n \in \{\pm 1\}$ represents the left ($-1$) or right ($+1$) shift at time step $n$. Here, for $n\in \{0,\,\cdots,\,N\}$ and $\bfit{\omega}\in \Omega_N$, we define the quantity $S_n = S_n^{(\bfit{\omega})}$ with $S_0^{(\bfit{\omega})} = 0$ and

\begin{align}\label{eq:sum_posi}
	S_n^{(\bfit{\omega})} = \sum_{t = 0}^{n-1} \omega_t\quad (n\geq 1).
\end{align}

Then $S_n$ represents the position of a walker at the time step $n$. Finally, we define the function $r: \{0,\,1,\,\cdots,\,N\}\to [0,\ 1]$ as

\begin{align}\label{eq:rn_def}
	r(n) := \mathbb{P}(\{\bfit{\omega}\in \Omega_N\,|\, \sgn(S_N^{(\bfit{\omega})}\cdot S_n^{(\bfit{\omega})}) = 1\}).
\end{align}

In this paper, we treat this as the figure-of-merit of future direction of QWRW or QSRW.

\appndx{Proof of Proposition~\ref{prp:rl}}{app:rl}\setcounter{equation}{0}
In the following, $\theta(s)$ is the function defined as Eq.~\refeq{eq:theta}\revise{, and let $F: [-|a|,\, |a|]\to \mathbb{C}$ be a continuous function. Then, we consider the following integral:
\begin{align}\label{eq:target_integral}
    I_n = \int_{-|a|}^{y}F(s)\exp(\pm 2in\theta(s))\rd s
\end{align}
with $y\in [-|a|,\, |a|]$. In fact, since $F(s)$ is uniformly continuous on the interval $[-|a|,\,|a|]$, it can be approximated by the following step function $F_J(s)$ with arbitrary precision, where $J$ is an appropriate natural number:
\begin{align}\label{eq:step}
    F_J(s) = \sum_{j=1}^{J}F(s_j)\mathbf{1}_{[s_{j-1},\, s_{j}]}(s).
\end{align}
Here $s_j$ for $j= 0,\ 1,\ \cdots,\ J$ is defined as follows:
\begin{align}\label{eq:s}
    s_j = -|a| + \fraction{2|a|j}{J}.
\end{align}
Besides, the function $\mathbf{1}_{[s_{j-1},\, s_{j}]}(s)$ is defined by
\begin{align}\label{eq:indicator_interval}
    \mathbf{1}_{[s_{j-1},\, s_{j}]}(s) = \left\{\begin{array}{ll} 1 & (s\in{[s_{j-1},\, s_{j}]}) \\ 0 & (s\not \in{[s_{j-1},\, s_{j}]})\end{array}\right..
\end{align}
It means that $s_j$-s are equispaced points with $-|a| = s_0 < s_1 < ... < s_J = |a|$. Therefore, it is sufficient to consider the integral $I_n^{(J)}$, which is made by replacing $F$ with $F_J$. That is, $I_n^{(J)}$ is described as
\begin{align}\label{eq:integral_step1}
    I_n^{(J)} = \int_{-|a|}^{y}F_J(s)\exp(\pm 2in\theta(s))\rd s.
\end{align}
By Eq.~\refeq{eq:step}, this can be rewritten as
\begin{align}\label{eq:integral_step2}\begin{split}
    I_n^{(J)} = \sum_{j=1}^{j_0}F(s_j)\int_{s_{j-1}}^{s_{j}}\exp(\pm 2in\theta(s))\rd s+ F(s_{j_0+1})\int_{s_{j_0}}^{y}\exp(\pm 2in\theta(s))\rd s,
\end{split}\end{align}
where $j_0$ satisfies $s_{j_0}\leq y\leq s_{j_0+1}$.}
	
\revise{Here w}e introduce the following lemma:
\revise{
\begin{lem}\label{lem:rl0}
    For any $\ell$ and $r$ that satisfy $-|a|\leq \ell< r\leq |a|$, it follows that
    \begin{align}\label{eq:rl0}
    \lim_{n\to\infty} \int_\ell^r \exp(\pm 2in\theta(s))\rd s =0.
    \end{align}
\end{lem}
}

\revise{Using this lemma and Eq.~\refeq{eq:integral_step2}, we obtain
\begin{align}\label{eq:integral_step_converge}
	\lim_{n\to\infty}I_n^{(J)} = 0.
\end{align}
This result leads to
\begin{align}\label{eq:integral_converge}
	\lim_{n\to\infty}I_n = 0.
\end{align}
}
Since the relations 
\begin{align}
    \cos(2n\theta(s)) &= \fraction{1}{2}(\exp(2in\theta(s))+\exp(-2in\theta(s))),\label{eq:cos_exp}\\ \sin(2n\theta(s)) &= \fraction{1}{2i}(\exp(2in\theta(s))-\exp(-2in\theta(s)))\label{eq:sin_exp}
\end{align}
hold, our proposition is obtained.

\subsection*{Proof of Lemma~\ref{lem:rl0}}
\revise{First, we consider the case that the interval $[\ell,\,r]$ does not include any of $\{0,\,\pm |a|\}$.}

We set $t = k(s)$, where $k(s)$ is given by Eq.~\refeq{eq:k}. Then, we obtain
\begin{align}\label{eq:sigma}
	s = \fraction{\sin t}{\sqrt{|b/a|^2 +\sin^2 t}} =: \sigma (t).
\end{align}
Note that $\sigma: [-\pi/2,\,\pi/2]\to [-|a|,\,|a|]$ is odd and \revise{strictly monotonically increasing, and $\sigma^{-1} = k$}. In addition, we can rewrite $\theta(s)$, given by Eq.~\refeq{eq:theta}, as
\begin{align}\label{eq:rho}
	\theta(s) = \arccos(|a|\cos t) -t \sigma(t) =: \fraction{1}{2}\rho(t).
\end{align}
This function $\rho(t) = 2\theta(s)$ is symmetric about $t =0$. Moreover, its derivative $\rho'(t)$ is positive for $t\in (-\pi/2,\,0)$ and zero on $t\in \{0,\, \pm\pi/2\}$.

\revise{Incidentally, the case we are considering is equivalent to $-|a|<\ell <r <0$ or $0<\ell <r <|a|$. By the symmetry of $\rho(t)$ and $\sigma(t)$, we can focus on the former case; the discussion on the latter case is made similarly. In the following, we consider the function $\tilde{\rho} = \rho|_{[-\pi/2,\,0]}$, which is the restriction map of $\rho$ to $[-\pi/2,\,0]$. Remark that $\tilde{\rho}$ is strictly monotonically increasing.}

Here, we rewrite the integration on the left-hand side of Eq.~\refeq{eq:rl0} with $\sigma(t)$ and $\revise{\tilde{\rho}}(t)$ as
\begin{align}\label{eq:int1}
	\int_\ell^r \exp(\revise{\pm}2in\theta(s))\rd s = \int_{k(\ell)}^{k(r)} \exp(\revise{\pm}in\revise{\tilde{\rho}}(t)) \sigma'(t)\rd t.
\end{align}
Moreover, making the substitution $\varPhi = \revise{\tilde{\rho}} (t)$, we have
\begin{align}\label{eq:int2}
	\int_{k(\ell)}^{k(r)} \exp(\revise{\pm}in\revise{\tilde{\rho}}(t)) \sigma'(t)\rd t =\int_{\revise{\tilde{\rho}\circ k}(\ell)}^{\revise{\tilde{\rho}\circ k}(r)} \exp(\revise{\pm}in\varPhi) \fraction{\sigma'\revise{\mathrel{\circ}\tilde{\rho}}^{-1}(\varPhi)}{\revise{\tilde{\rho}'\circ\tilde{\rho}}^{-1}(\varPhi)}\rd \varPhi.
\end{align}
Here, \revise{since the inequality $-{\pi}/{2}< \revise{k}(\ell)< \revise{k}(r) <0$ holds,} $\revise{\tilde{\rho}}'(t)>0$ holds on the interval $[\revise{k}(\ell),\,\revise{k}(r)]$. Therefore, the function \revise{${\sigma'\circ\tilde{\rho}^{-1}(\varPhi)}/{\tilde{\rho}'\circ\tilde{\rho}^{-1}(\varPhi)}$ on Eq.~\refeq{eq:int2}} is continuous on this interval, and thus we can apply the Riemann-Lebesgue lemma to the right-hand side of Eq.~\refeq{eq:int2}. That is,
\begin{align}\label{eq:int3}
	\lim_{n\to \infty}\int_{\revise{\tilde{\rho}\circ k}(\ell)}^{\revise{\tilde{\rho}\circ k}(r)} \exp(\revise{\pm}in\varPhi) \fraction{\sigma'\revise{\mathrel{\circ}\tilde{\rho}}^{-1}(\varPhi)}{\revise{\tilde{\rho}'\circ\tilde{\rho}}^{-1}(\varPhi)}\rd \varPhi =0.
\end{align}

Summarizing the discussion above, we obtain the desired result \revise{in the case that the relation $-|a|<\ell <r <0$ holds. Once again, the case that $0<\ell <r <|a|$ can be discussed in a similar manner.}

\revise{In the following, we discuss the general case.} We take $\varepsilon$ that satisfies $0<\varepsilon <r-\ell$. \revise{Besides, we define the open intervals $O_1,\ O_2,$ and $O_3$ as follows:
\begin{align}
	O_1 := \left(-|a|-\fraction{\varepsilon}{6},\ -|a|+\fraction{\varepsilon}{6}\right), \quad
	O_2 := \left(-\fraction{\varepsilon}{6},\ \fraction{\varepsilon}{6}\right), \quad
	O_3 := \left(|a|-\fraction{\varepsilon}{6},\ |a|+\fraction{\varepsilon}{6}\right). \label{eq:o3}
\end{align}
Note that putting $\mathbb{U} = O_1\cup O_2\cup O_3$, the Lebesgue measure of $\mathbb{U}$ is $\mathcal{L}(\mathbb{U}) = \varepsilon$, and $\{0,\,\pm |a|\}$ is included in $\mathbb{U}$. That is, the set $[\ell,\,r]\setminus \mathbb{U}$ is a closed interval that does not contain any of $\{0,\,\pm|a|\}$, or the sum of such ones. Therefore, by the restricted result above, we obtain
\begin{align}\label{eq:lim0}
	\lim_{n\to \infty}\int_{[\ell,\,r]\backslash\mathbb{U}}\exp(\pm 2in\theta(s))\rd s =0.
\end{align}
}

\revise{
By the triangle inequality, it holds that
\begin{align}\label{eq:triangle}\begin{split}
    &\left|\int_{[\ell,\,r]}\exp(\pm 2in\theta(s))\rd s\right| \\
    &\leq \left|\int_{[\ell,\,r]}\exp(\pm 2in\theta(s))\rd s -\int_{[\ell,\,r]\backslash\mathbb{U}}\exp(\pm 2in\theta(s))\rd s\right|+\left|\int_{[\ell,\,r]\backslash\mathbb{U}}\exp(\pm 2in\theta(s))\rd s\right|.
\end{split}\end{align}
The first term of the right-hand-side in this inequality can be estimated by $\varepsilon$ as follows:}
\begin{align}\label{eq:estimate}
	\left| \int_{[\ell,\,r]}\exp(\revise{\pm} 2in\theta(s))\rd s -\int_{[\ell,\,r]\backslash\mathbb{U}}\exp(\revise{\pm} 2in\theta(s))\rd s\right|\leq \int_{[\ell,\,r]\backslash([\ell,\,r]\backslash\mathbb{U})}|\exp(\revise{\pm} 2in\theta(s))|\rd s \leq \int_{\mathbb{U}}\rd s = \varepsilon.
\end{align}
\revise{Applying Eqs.~\refeq{eq:lim0} and \refeq{eq:estimate} to Eq.~\refeq{eq:triangle},} we obtain
\begin{align}
	\limsup_{n\to \infty}\left| \int_{[\ell,\,r]}\exp(\revise{\pm} 2in\theta(s))\rd s\right| \leq\varepsilon.
\end{align}

Since $\varepsilon$ can be taken to be as close to $0$ as one wants, the lemma follows. 

\appndx{Discussions on Oscillations of $\bfit{p_n(x)}$}{app:oscillation}\setcounter{equation}{0}
\def\parc{\mathrm{(c)}}
\begin{figure}[t]
\begin{minipage}[b]{0.5\linewidth}
	\centering\includegraphics[width=\textwidth]{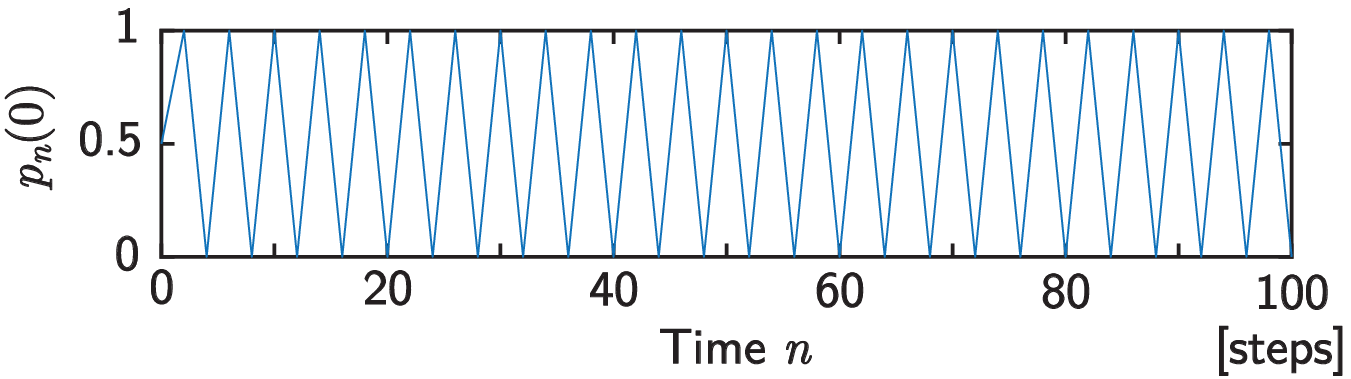}
	\subcaption{$a = 1/\sqrt{2}$}
\end{minipage}
\begin{minipage}[b]{0.5\linewidth}
	\centering\includegraphics[width=\textwidth]{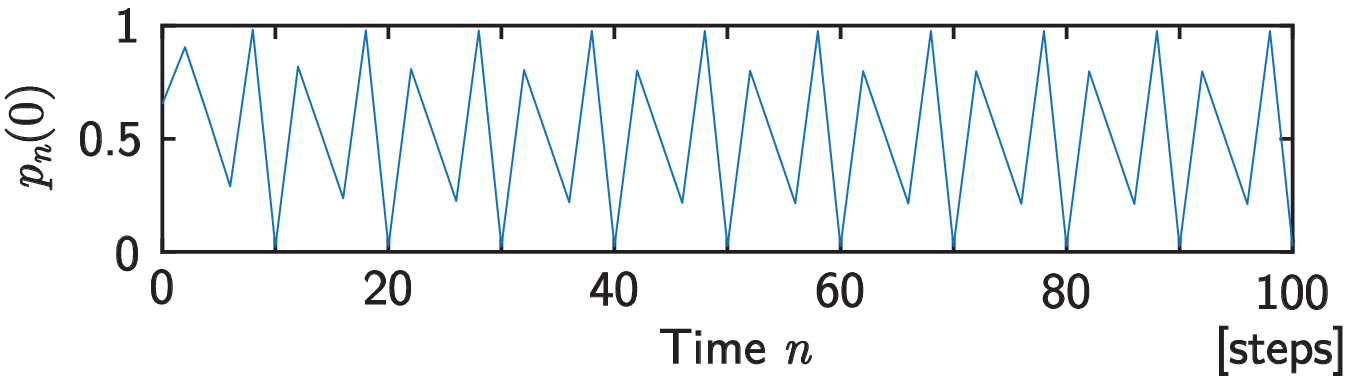}
	\subcaption{$a = \cos(\pi/5)$}
\end{minipage}
\caption{Graphs of $p_n(0)$ over time step $n$. Plots are given only to even $n$; if $n$ is odd, then $p_n(0) = 1/2$ by the definition \refeq{eq:p}. The initial state $\ket{\varphi} = [1\,\,\,0]^\trp$ for both cases.}\label{fig:oscillation}
\vrule height 0.3mm width 165mm\vspace{0.4\baselineskip}

\begin{minipage}[b]{\linewidth}
	\centering\includegraphics[width=\textwidth]{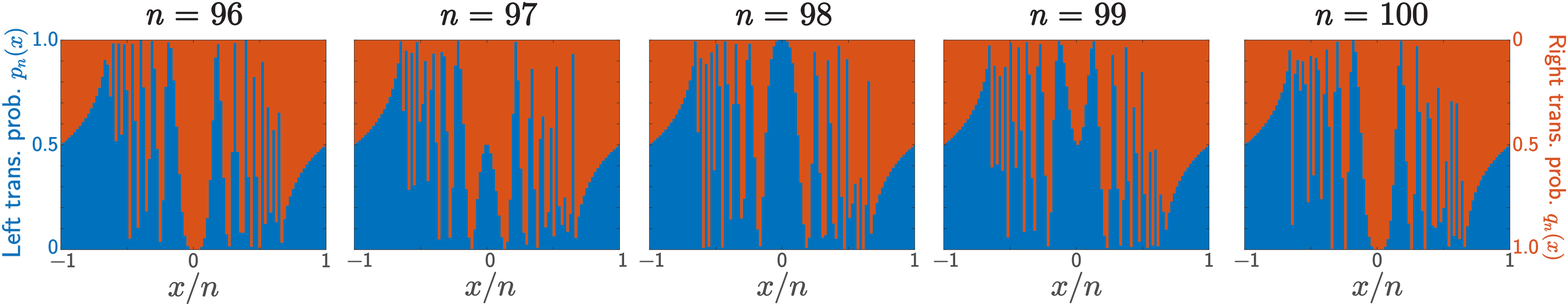}
	\caption{Time variation of transition probabilities $p_n(x)$ and $q_n(x)$ in the case of Fig.~\ref{fig:oscillation}. The values are expressed only on the even (resp. odd) points when the time step is even (resp. odd).}\label{fig:oscillation2}
	\vrule height 0.3mm width 165mm
\end{minipage}
\end{figure}

\revise{We define the function $\gamma: [0,\ \infty)\times (-1/\sqrt{2},\ 1/\sqrt{2})\to [0,\ 1]$ as
\begin{align}\label{eq:gamma}
	\gamma(t,\,s) = \fraction{\tau_1(s) +\tilde{\xi}^\parc(t,\,s)/\lambda(s)}{1 +\tilde{\eta}^\parc (t,\,s)/\lambda(s)}.
\end{align}
Here the functions $\tilde{\xi}^\parc,\ \tilde{\eta}^\parc : [0,\ \infty)\times (-1/\sqrt{2},\ 1/\sqrt{2})\to [0,\ 1]$ are given by
\begin{align}
	\tilde{\xi}^\parc (t,\,s) &:= -2\im \left( e^{2it\theta(s)}f_{\tilde{\varphi}}^{\rL}(s)\overline{g_{\revise{\tilde{\varphi}}}^{\rL}(s)}\right), \label{eq:xic}\\
	\tilde{\eta}^\parc (t,\,s) &:= -2 \im \left(e^{2it\theta(s)}(f_{\revise{\tilde{\varphi}}}^{\tilde{\rL}}(s)\overline{g_{\revise{\tilde{\varphi}}}^{\tilde{\rL}}(s)} +f_{\revise{\tilde{\varphi}}}^{\tilde{\rR}}(s)\overline{g_{\revise{\tilde{\varphi}}}^{\tilde{\rR}}(s)})\right), \label{eq:etac}
\end{align}
and $\tau_1(s)$, $\lambda(s)$, $f^\psi_\chi(s)$ and $g^\psi_\chi(s)$ in Eqs.~\refeq{eq:gamma}, \refeq{eq:xic}, and \refeq{eq:etac} are given by Eqs.~\refeq{eq:taus}, \refeq{eq:lambdas}, \refeq{eq:fpsi}, and \refeq{eq:gpsi}, respectively. It is remarkable that functions $\tilde{\xi}_n(s)$ and $\tilde{\eta}_n(s)$ in Eqs.~\refeq{eq:xitilde} and \refeq{eq:etatilde} are the restricted functions of $\tilde{\xi}^\parc(t,\,s)$ and $\tilde{\eta}^\parc(t,\,s)$ to $\mathbb{N}_0\times (-1/\sqrt{2},\ 1/\sqrt{2})$, and thus the relation
\begin{align}\label{eq:p_gamma}
	p_n(x_n) = \gamma\left(n,\, \fraction{x_n}{n}  + O\left(\fraction{1}{n^2}\right)\right)
\end{align}
holds by Eq.~\refeq{eq:pnxn}.}

\revise{The function $\gamma(t,\,s)$ has both temporal ($t$) and spatial ($s$) oscillations resulting from the trigonometric functions included in $\tilde{\xi}^\parc(t,\,s)$ and $\tilde{\eta}^\parc(t,\,s)$. These oscillations in turn create the oscillatory behavior of $p_n(x)$.}

\revise{First, we shall discuss the temporal oscillations. 
For a fixed $s\in (-1/\sqrt{2},\ 1/\sqrt{2})$, the temporal oscillation of $\gamma(t,\,s)$ has the periodicity $T(s) = \pi/\theta(s)$, where $\theta$ is given by Eq.~\refeq{eq:theta}. 
The temporal oscillations of $p_n(x)$ is strongly affected by $\gamma(t,\,s)$ and show the similar periodic behavior to it in some cases. Let $x$ fix to $0$. Then, if $n$ is even, $p_n(0)$ is described as
\begin{align}\label{eq:pn0_gamma}
	p_n(0) = \gamma \left(n,\, O\left(\fraction{1}{n^2}\right)\right).
\end{align}
This indicates that $p_n(0)$ is approximated by $\gamma (n,\,0)$ for sufficiently large $n$ if $n$ is even. Therefore, considering $p_n(0)$ is $1/2$ if $n$ is odd, and $T(0) = \arccos(|a|)$ holds by $\theta(0) = \pi/ \arccos(|a|)$, we can classify the periodicity $\mathcal{T}$ of $p_n(0)$ for sufficiently large $n$ as follows:
\begin{prp}\label{prp:periodicity}
Assume that there exists a pair of the relatively prime natural numbers $(\ell,\,m)$ that satisfies $0\leq \ell < 2m$ and $a = \cos (\ell/m)\pi$. Then,
\begin{align}
	\mathcal{T} = \left\{ \begin{array}{cl} \ell m & (\text{if $\ell$ is even}) \vrule width 0pt height 0pt depth 10pt\\
	 2\ell m & (\text{if $\ell$ is odd}) \end{array}\right. .
\end{align}
If this assumption is not true, then $\mathcal{T} = \infty$; that is, $p_n(0)$ is aperiodic.
\end{prp}}

\revise{Figs.~\ref{fig:oscillation}(a) and (b) show the value of $p_n(0)$ in the case of $a= 1/\sqrt{2}$ and $a = \cos(\pi/5)$, respectively. For both cases, the initial state $\ket{\varphi}$ is set to be $[1\,\,\,0]^\trp$. In the (a)-case, the corresponding $\gamma(t,\,s)$ has periodicity $T(0) = 4$, and we can observe the same one ($\mathcal{T} = 4$) from $p_n(0)$. In the (b)-case, the corresponding $\gamma(t,\,s)$ has periodicity $T(0) = 5$, and we can observe periodicity $\mathcal{T} = 10$ from $p_n(0)$. These facts regarding the relation between $T(0)$ and $\mathcal{T}$ agree with Proposition~\ref{prp:periodicity}. Note that the behavior of $p_n(x)$ is unstable at early $n$ due to the error term $O(1/n^2)$, but the periodic behavior is clearly observed for sufficiently large $n$. }

\revise{Fig.~\ref{fig:oscillation2} shows time variation of $p_n(x)$ and $q_n(x)$ in the case of Fig.~\ref{fig:oscillation}(a); i.e., $a = 1/\sqrt{2}$. Herein, the periodic behavior is observed, which is investigated in Fig.~\ref{fig:oscillation}(a). Incidentally, $\gamma(t,\,0)$ in this case is calculated as
\begin{align}\label{eq:oscillation_a_case}
	\gamma(t,\,0) = \fraction{1}{2} \left(1-\cos \left(\fraction{\pi}{2}t\right)\right),
\end{align}
and this fact seems to affect the behavior near the origin, the clarification of which is still open.}

\begin{figure}[t]
    \centering\includegraphics[width=0.5\textwidth]{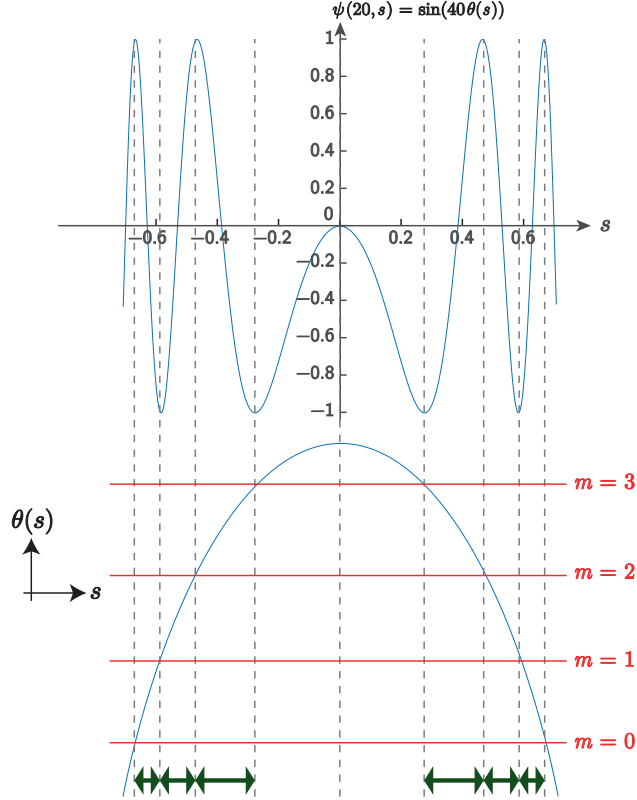}
	\caption{\revise{Functions $\psi(t,\,s) = \sin(2t\theta(s))$ and $\theta(s)$ ($t=20$).}}\label{fig:psins}
\vrule height 0.3mm width 165mm
\end{figure}

\revise{Second, we discuss the spatial oscillation. As shown in Fig.~\ref{fig:pq}, $p_n(x)$ has high-frequency spatial oscillation, and the frequency is higher for positions farther from the origin. Our desire is to explain this phenomenon via $\gamma(t,\,s)$, but a full mathematical analysis has not been made yet. 
However, a simple argument based on the function $\psi(t,\,s) = \sin(2t\theta(s))$ can be made. The function $\psi(t,\,s)$ has in common with $\gamma(t,\,s)$ that the farther from the origin one goes, the higher the frequency of oscillation becomes, see Fig.~\ref{fig:psins}. We can interpret this phenomenon with the intervals of extrema of $\psi(t,\,s)$; that is, the intervals of extrema get narrower if the absolute value of $s$ get larger. Assume that $t$ is fixed and satisfies $t>0$. If $\psi(t,\,s)$ has an extremum on a certain $s$, then the equation
\begin{align}\label{eq:deri_psi}
	\fraction{\partial \psi(t,\,s)}{\partial s} = 0
\end{align}
holds. This equation holds iff either ${\partial \theta(s)}/{\partial s} = 0$ or $\cos(2t\theta(s)) = 0$ holds. Here the former condition is equivalent to $s=0$ because $\theta(s)$ is concave. The latter condition is equivalent to that there exists $m\in \mathbb{N}_0$ such that the following holds:
\begin{align}\label{eq:theta_m}
	\theta(s) = \fraction{2m+1}{4t}\pi.
\end{align}
This indicates that $\theta(s)$ is taken in equidistant intervals, and the corresponding $s$ has extrema of $\psi(t,\,s)$. Here $\theta(s)$ is concave, and thus intervals of two extrema get narrower on the larger-$s$ area as visualized in Fig.~\ref{fig:psins}. Therefore the frequency of oscillations of $\psi(t,\,s)$ is higher on the area farther from $s=0$. Whether the similar discussion can be made for $\gamma(t,\,s)$ is an open question at this point.}


\vspace{\baselineskip}

\noindent
\textbf{Acknowledgments:} 
This work was supported by the SPRING program (JPMJSP2108), the CREST project (JPMJCR17N2) funded by the Japan Science and Technology Agency, and Grants-in-Aid for Scientific Research (JP20H00233) and Transformative Research Areas (A) (JP22H05197) funded by the Japan Society for the Promotion of Science. A. R. is a JSPS International Research Fellow.

\vspace{\baselineskip}\noindent
\textbf{Data availability statement:} The datasets generated during and/or analysed during the current study are available from the corresponding author on reasonable request.

\bibliographystyle{ieeetr}
\bibliography{skeleton_Bib}
\end{document}